\documentclass[reqno,nofootinbib,reprint]{revtex4-1}
\usepackage{tikz}[2010/10/13]
\usepackage{breakurl}
\usepackage{ifpdf}
\input{Qcircuit}
\usepackage{mathtools}
\usepackage{hyperref}
\usepackage{amssymb,color}

\newtheorem{Thm}{Theorem}[section]

\newtheorem{remark}[Thm]{Remark}

\newcommand{\abs}[1]{\left\vert #1 \right\vert}
\DeclareMathOperator{\Tr}{tr}
\newcommand{\be}{\begin{equation}}
\newcommand{\ee}{\end{equation}}
\newcommand{\blue}[1]{{\color{blue}#1}}

\begin{document}
\title[Qudit Isotopy]{Qudit Isotopy}
\author{Arthur Jaffe}
\email{arthur\_jaffe@harvard.edu}
\author{Zhengwei Liu}
\email{zhengweiliu@fas.harvard.edu}
\author{Alex Wozniakowski}
\email{airwozz@gmail.com}
\affiliation{Harvard University, Cambridge, MA 02138, USA}
\begin{abstract}
We explore a general diagrammatic framework to understand qudits and their braiding, especially in its relation to entanglement. This involves understanding the role of isotopy in interpreting diagrams that implement entangling gates as well as some standard quantum information protocols.  We give qudit Pauli operators $X,Y,Z$ and comment on their structure, both from an algebraic and from a diagrammatic point of view.  We explain alternative models for diagrammatic interpretations of qudits and their transformations. We use our diagrammatic approach to define an entanglement-relay protocol for long-distance entanglement. Our approach rests on algebraic and topological relations discovered in the study of planar para algebras. In summary, this work provides bridges between the new theory of planar para algebras and quantum information, especially in questions involving entanglement.
\end{abstract}
\maketitle

\section{Introduction}
In this paper we give various diagrammatic models of qudits. In our first model, we represent one qudit as a string; in the second (two-string) model it becomes a cap; in a third (four-string) model it is represented by a pair of caps. The particles in each of these models may be parafermions, or bosons that arise as parafermionic particle-anti-particle pairs. 

It is the robust nature of these diagrams, which illustrate properties of states and of operators, that fascinates us.
Our general approach is an application of the mathematical framework of \textit{planar para algebras}, that we introduced in~\cite{JL}.  In that paper we elaborate on the general topological properties of the models that we only sketch here. 

The main goal of our present work is to provide a link between the new theory of planar para algebras and quantum information.   
We give a solution to the Yang-Baxter equation that represents a braid. In the first two models, the braiding of qudits describes  qudit entanglement. We also use braiding of qudits to obtain the entanglement distribution protocol, the entanglement-swapping protocol and the entanglement-relay protocol.

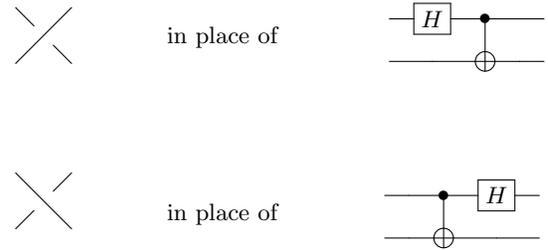
\begin{figure}[h]
\centering
%\begin{subfigure}{0.5\textwidth}
\begin{tikzpicture}
\draw (1.25,0.5) --(1,0.75);
\draw (1.75,0) --(1.5,0.25);
\draw (1,0) --(1.75,0.75);
\node  (3.75,0.35) at (3.75,0.35) {in place of};
\node  (7,0.35) at (7,0.35) {\Qcircuit @C=1em @R=.7em {
& \gate{H} & \ctrl{1} & \qw & \qw \\
& \qw & \targ & \qw &\qw }};
\draw (1.75,-2.2) --(1,-1.45);
\draw (1,-2.2) --(1.25,-1.95);
\draw (1.5,-1.7) --(1.75,-1.45);
\node  (3.75,0.35) at (3.75,-2) {in place of};
\node  (7,0.35) at (7.1,-2) {\Qcircuit @C=1em @R=.7em {
& \qw & \ctrl{1} & \gate{H} & \qw & \\
& \qw & \targ & \qw & \qw }}; \end{tikzpicture}
%\end{subfigure} 
\centering
\caption{Entangling, unitary solution of the Yang-Baxter equation on the left, and entangling quantum circuit on the right.
\label{Braid-for-Gate}
}
\end{figure}

We illustrate in Figure \ref{Braid-for-Gate} the use of a braid to simulate the entanglement provided by the conjunction of a Hadamard and CNOT gate.
We construct a similar maximally entangling qudit-braid.

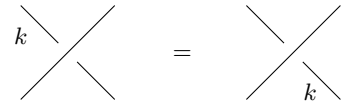
\begin{figure}[h]
\centering
%\begin{subfigure}{0.5\textwidth}
\begin{tikzpicture}
\node  (3.75,0.35) at (1,0.85) {$k$};
\draw (1.5,0.75) --(1,1.25);
\draw (2.25,0) --(1.75,0.5);
\draw (1,0) --(2.25,1.25);
\node  (3.75,0.35) at (3.15,0.6) {{$=$}};
\draw (4.5,0.75) --(4,1.25);
\draw (5.25,0) --(4.75,0.5);
\draw (4,0) --(5.25,1.25);
\node  (3.75,0.35) at (4.85,0.1) {$k$};
\end{tikzpicture}
%\end{subfigure} 
\centering
\caption{Qudit-braid relation.
\label{Braid-qudit}
}
\end{figure}

We use braids such as in Figure \ref{Braid-for-Gate}, but generalized to include particle excitations illustrated in Figure \ref{Braid-qudit}. Here the particle with charge $k$ is represented by the label $k$.
As a consequence of the Brylinskis' remarkable criterion, one can employ this braid to obtain a partial topological quantum computer  for parafermions.\footnote{This criterion is Theorem 4.1 of \cite{B}. The preprint and published versions have different organization, and we refer to the numbering in the latter.  See also \cite{two-qubit gate}.}

The \textit{qudit-braid} relation illustrated in Figure \ref{Braid-qudit} shows how a particle moves under the braid crossing. This identity allows us to use topological isotopy in three-dimensional space.  This technique was used before in planar algebras, but it is new in the context of braids with particle excitations.

We explain our notation in \S\ref{Notation}, including the  interpretation of the structure of states in terms of diagrams, as well as the interpretation of the trace and partial trace--which enter the process of measurement. In \S\ref{Particles} we introduce braids that involves particle excitations.  There we explain the qudit-braid relation.   

In \S\ref{XYZ Matrices} we focus on two different versions of qudit Pauli  $X,Y,Z$ matrices, which are useful for interpreting protocols. The diagrammatic presentation of these matrices makes clear the way $X,Y,Z$ are built from qudits, and how one can translate the qudit representation into formulas. In particular, in our four-string model one sees from the diagrams how and why the matrices $X,Y,Z$ act on the charge-zero (gauge-invariant) subspace of a space of qudits.    

In \S\ref{Protocols} we give some applications of the diagrammatic method to understanding entanglement protocols.  We address the entanglement-distribution protocol and the entanglement-swapping protocol.   We go into one application in detail, in which we realize a quantum circuit using the one-string model (that we employ throughout the bulk of the paper). This model illustrates how we take advantage of  \textit{topological isotopy}---a property central to the structure of planar para algebras. 

In \S\ref{Sect:DifferentModels-1}--\S\ref{Sect:DifferentModels-4} we contrast our one-string, two-string, and four-string models.  (The four-string model is especially adaptable to certain situations with redundant degrees of freedom, including models for $X,Y,Z$.  Here charge neutrality of qudits as particle-anti-particle pairs plays a natural role.)

In \S\ref{Sect:ControlledGates} we discuss some further applications. In particular we show how our four-string model easily describes controlled gates, that have been studied algebraically in a recent paper of Hutter and Loss~\cite{HutterLoss}.

In \S\ref{Sect:ERelay} we define an \textit{entanglement-relay}  protocol to implement long-distance entanglement. This protocol allows one to transfer entanglement in a non-local fashion to distant objects. 

\section{Notation}\label{Notation}
\subsection{The Parafermion Algebra}
The \textit{parafermion algebra} is a $\ast$-algebra with unitary generators $c_{j}$, which satisfy
\begin{equation}\label{ParafermionAlgebra}
c_{j}^{d}=1 \; \; \; \mbox{and} \; \; \; c_{j}c_{k}=q \, c_{k}c_{j} \; \; \; \mbox{for} \; \; 1 \leqslant  j< k \leqslant m.
\end{equation}
\noindent Here $q \equiv e^{\frac{2 \pi i}{d}}$, $i \equiv \sqrt{-1}$, and $d$ is the order of the parafermion.
Consequently $c^{\ast}_{j}=c^{-1}_{j}=c^{d-1}_{j}$ where $\mbox{*}$ denotes the adjoint. Majorana fermions arise for $d=2$. This is an example of a planar para algebra, for which the general theory provides diagrammatic representations:  for elements of this algebra, and for the representation of its action on Hilbert space.

\subsection{Diagrammatic Representation} \label{hilbert space}

We introduce diagrams to represent elements of our algebra or qudits. The diagrams multiply from bottom to top\footnote{This follows standard conventions for braids, while the standard convention for circuits is multiplication from left to right.}. Also, tensor products multiply from left to right. We represent the horizontal multiplication $AB$ and the tensor product $A \otimes B$ by
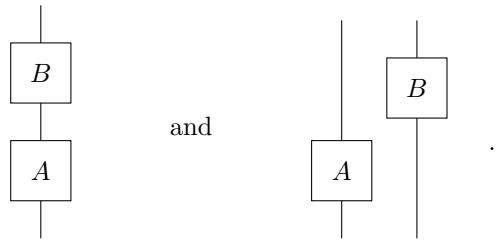
\begin{center}
\begin{tikzpicture}
\draw (0,0) --(0,0.5);
\draw (-0.4,.5) rectangle +(0.8,0.8);
\node (0,0.85) at (0,0.9) {$A$};
\draw (0,1.3) --(0,1.8);
\draw (-0.4,1.8) rectangle +(0.8,0.8);
\node (0,2.2) at (0.0,2.2) {$B$};
\draw (0,2.6) --(0,3.1);
\node (2,0) at (2,1.5) {$\mbox{and}$};
\draw (4,0) --(4,0.5);
\draw (3.6,.5) rectangle +(0.8,0.8);
\node (4,0.85) at (4,0.9) {$A$};
\draw (4,1.3) --(4,2.9);
\draw (5,0) --(5,1.6);
\draw (4.6,1.6) rectangle +(0.8,0.8);
\node (0,0.85) at (5.0,2) {$B$};
\draw (5,2.4) --(5,2.9);
\node (2,0) at (6,1.2) {.};
\end{tikzpicture}\end{center}
\noindent We represent a generator $c_{j}$ in the $j^{\mbox{th}}$ tensor factor as
\begin{center}
\begin{tikzpicture}
\node (-0.5,0.25) at (-1.6,0.5) {$c_{j} \; \; \mbox{replaced by}$};
\draw (0,0) --(0,1);
\draw (0.2,0) --(0.2,1);
\draw [fill] (0.4,0.5) circle [radius=0.01];
\draw [fill] (0.5,0.5) circle [radius=0.01];
\draw [fill] (0.6,0.5) circle [radius=0.01];
\draw (0.95,0) --(0.95,1);
\node (0,0) at (0.85,0.5) {$1$};
\node (0.9,0) at (0.95,-0.25) {$j$};
\draw [fill] (1.2,0.5) circle [radius=0.01];
\draw [fill] (1.3,0.5) circle [radius=0.01];
\draw [fill] (1.4,0.5) circle [radius=0.01];
\draw (1.6,0) --(1.6,1);
\draw (1.8,0) --(1.8,1);
\node (2,0) at (2.2,0.4) {.};
\end{tikzpicture} \end{center}
\noindent The power $c_{j}^{\alpha}$ of $c_{j}$ arises from replacing the label ``$1$'' by the label ``$\alpha$.'' Additionally,
\begin{center}
\begin{tikzpicture}
\draw (0,0) --(0,1);
\node (0,0) at (-0.2,0.8) {$\beta$};
\node (0,0) at (-0.2,0.2) {$\alpha$};
\node (0.45,0.25) at (0.45,0.5) {$=$};
\node (0.9,0.35) at (1.4,0.5) {$\alpha + \beta$};
\draw (2,0) --(2,1);
\node (1.4,-0.05) at (2.3,0.3) {,};
\node (2.2,0.35) at (3.4,0.4) {$\mbox{and}$};
\node (3.7,0.35) at (4.8,0.5) {$d$};
\draw (5,0) --(5,1);
\node (4.4,0.25) at (5.4,0.5) {$=$};
\draw (5.8,0) --(5.8,1);
\node (4.95,0) at (6.1,0.3) {.};
\end{tikzpicture} \end{center}

\noindent The parafermion relation \eqref{ParafermionAlgebra} becomes
\begin{equation}
\begin{tikzpicture}
\draw (0,0) --(0,01);
\draw (0.2,0) --(0.2,01);
\draw [fill] (0.4,0.5) circle [radius=0.01];
\draw [fill] (0.5,0.5) circle [radius=0.01];
\draw [fill] (0.6,0.5) circle [radius=0.01];
\node (0,0) at (0,-0.25) {$j$};
\node (0,0) at (-0.15,0.2) {$\alpha$};
\draw (0.8,0) --(0.8,1);
\draw (1.15,0) --(1.15,1);
\node (0.3,0) at (1.15,-0.25) {$k$};
\node (1.2,0) at (1,0.8) {$\beta$};
\node (0.9,0.15) at (1.6,0.35) {$=$};
\node (0.9,0.15) at (2.2,0.4) {$q^{{\alpha}{\beta}}$};
\draw (2.9,0) --(2.9,1);
\draw (3.1,0) --(3.1,1);
\draw [fill] (3.3,0.5) circle [radius=0.01];
\draw [fill] (3.4,0.5) circle [radius=0.01];
\draw [fill] (3.5,0.5) circle [radius=0.01];
\node (2.4,0) at (2.9,-0.25) {$j$};
\node (0,0) at (2.75,0.8) {$\alpha$};
\draw (3.7,0) --(3.7,1);
\draw (4.05,0) --(4.05,1);
\node (1.8,0) at (4.05,-0.25) {$k$};
\node (0,0) at (3.9,0.2) {$\beta$};
\node (3.3,0.1) at (5.8,0.3) {$\mbox{for} \; \, j < k$};
\node (4.2,0) at (6.8,0.2) {,};
\end{tikzpicture}
\label{ParafermionRelationDiagram}
\end{equation}
where the strings between $j$ and $k$ contain no excitations. We call $q^{\alpha \beta}$ the twisting scalar. 

Let $\zeta$ be a square root of $q$, with the property $\zeta^{d^{2}}=1$.
We remark that the diagrammatic interpretation given in \cite{JL} of the twisted tensor product $X\circ Y=\zeta^{\abs{X}\abs{Y}}\,XY$ introduced in \cite{JP,JJ}, interpolates between the left and right side of the parafermion relation \eqref{ParafermionRelationDiagram}.  We write the labels on the same vertical height. Then
\begin{equation}
\begin{tikzpicture}
\draw (0,0) --(0,01);
\draw (0.2,0) --(0.2,01);
\draw [fill] (0.4,0.5) circle [radius=0.01];
\draw [fill] (0.5,0.5) circle [radius=0.01];
\draw [fill] (0.6,0.5) circle [radius=0.01];
\node (0,0) at (0,-0.25) {$j$};
\node (0,0) at (-0.15,0.2) {$\alpha$};
\draw (0.8,0) --(0.8,1);
\draw (1.15,0) --(1.15,1);
\node (0.3,0) at (1.15,-0.25) {$k$};
\node (1.2,0) at (1,0.8) {$\beta$};
\node (0.9,0.15) at (1.6,0.35) {$=$};
\node (0.9,0.15) at (2.2,0.4) {$\zeta^{{\alpha}{\beta}}$};
\draw (2.9,0) --(2.9,1);
\draw (3.1,0) --(3.1,1);
\draw [fill] (3.3,0.5) circle [radius=0.01];
\draw [fill] (3.4,0.5) circle [radius=0.01];
\draw [fill] (3.5,0.5) circle [radius=0.01];
\node (2.4,0) at (2.9,-0.25) {$j$};
\node (0,0) at (2.75,0.5) {$\alpha$};
\draw (3.7,0) --(3.7,1);
\draw (4.05,0) --(4.05,1);
\node (1.8,0) at (4.05,-0.25) {$k$};
\node (0,0) at (3.9,0.5) {$\beta$};
\node (3.3,0.1) at (5.8,0.3) {$\mbox{for} \; \, j < k$};
\node (4.2,0) at (6.8,0.2) {.};
\end{tikzpicture} \end{equation}

The diagram called a \textit{cap} is not an element of the parafermion algebra. Rather it is a vector that provides one qudit.  We transport the qudit label from left to right on the cap, producing a phase $\zeta$, which can be interpreted as a Fourier transform relation, see \cite{JL}.  The cap has the form 
\begin{center}
\begin{tikzpicture}
\node (2.9,0.6) at (-0.9,-0.3) {$\alpha$};
\coordinate (A) at (-0.3,0);
\coordinate (B) at (0.3,0);
\draw[line width=0.25mm,color=black] (A) to [bend left=30] (B);
\coordinate (A) at (-0.6,-0.5);
\coordinate (B) at (-0.3,0);
\draw[line width=0.25mm,color=black] (A) to [bend left=30] (B);
\coordinate (A) at (0.6,-0.5);
\coordinate (B) at (0.3,0);
\draw[line width=0.25mm,color=black] (B) to [bend left=30] (A);
\node (2.9,0.6) at (1.1,-0.2) {$=$};
\coordinate (A) at (2.2,-0.5);
\coordinate (B) at (2.5,0);
\draw[line width=0.25mm,color=black] (A) to [bend left=30] (B);
\coordinate (A) at (2.5,0);
\coordinate (B) at (3.1,0);
\draw[line width=0.25mm,color=black] (A) to [bend left=30] (B);
\coordinate (A) at (3.4,-0.5);
\coordinate (B) at (3.1,0);
\draw[line width=0.25mm,color=black] (B) to [bend left=30] (A);
\node (2.9,0.6) at (1.7,-0.15) {$\zeta^{\alpha^{2}}$};
\node (2.9,0.6) at (3.15,-0.3) {$\alpha$};
\node (0,0) at (3.65,-0.25) {$.$};
\end{tikzpicture}\;\end{center}
\noindent

We represent the adjoint $\mbox{*}$ diagrammatically as
\begin{equation}
\begin{tikzpicture}
\node (-0.6,0.25) at (-0.7,0.5) {$\ast$};
\node (-0.3,0.25) at (-0.4,0.5) {$\colon$};
\draw (0,0) --(0,1);
\node (0,0) at (-0.1,0.5) {$1$};
\draw[->] (0.2,0.5)--(0.63,0.5);
\node (1.2,0.55) at (1.25,0.5) {$d$$-$$1$};
\draw (1.7,0) --(1.7,1);
\node (1.9,-0.05) at (1.9,0.3) {.};
\end{tikzpicture} \end{equation}
\noindent More generally, the adjoint $\mbox{*}$ of a product comes from its vertical reflection,
\begin{center}
\begin{tikzpicture}
\coordinate (A) at (-0.7,0.25);
\coordinate (B) at (-0.7,2.6);
\draw[line width=0.25mm,color=black] (A) to [bend left=40] (B);
\coordinate (C) at (0.7,0.25);
\coordinate (D) at (0.7,2.6);
\draw[line width=0.25mm,color=black] (C) to [bend left=-40] (D);
\node (2.9,0.6) at (1.2,2.6) {$\ast$};
\draw (-0.4,.25) rectangle +(0.8,0.8);
\node (0,0.65) at (0,0.65) {$A$};
\draw (-0.05,1.05) --(-0.05,1.4);
\draw (0.05,1.05) --(0.05,1.4);
\draw (-0.4,1.4) rectangle +(0.8,0.8);
\node (0,1.8) at (0.0,1.8) {$B$};
\draw (-0.15,2.2) --(-0.15,2.6);
\draw (-0.05,2.2) --(-0.05,2.6);
\draw (0.05,2.2) --(0.05,2.6);
\draw (0.15,2.2) --(0.15,2.6);
\node (2.9,0.6) at (2,1.3) {$=$};
\draw (3.1,1.8) rectangle +(0.8,0.8);
\node (0,0.65) at (3.55,2.2) {$A^{\ast}$};
\draw (3.45,1.45) --(3.45,1.8);
\draw (3.55,1.45) --(3.55,1.8);
\draw (3.1,0.65) rectangle +(0.8,0.8);
\node (0,0.65) at (3.57,1.05) {$B^{\ast}$};
\draw (3.35,0.25) --(3.35,0.65);
\draw (3.45,0.25) --(3.45,0.65);
\draw (3.65,0.25) --(3.65,0.65);
\draw (3.55,0.25) --(3.55,0.65);
\node (2,0) at (4.5,1.1) {.};
\end{tikzpicture}\end{center}

The \textit{cup} diagram is related to the cap above, and it also satisfies a parafermion relation.  We obtain the cup from the cap by the adjoint, followed by the substitution $\alpha\to-\alpha$. Thus
\begin{center}
\begin{tikzpicture}
\node (2.9,0.6) at (-0.85,0.25) {$\alpha$};
\coordinate (A) at (-0.3,0);
\coordinate (B) at (0.3,0);
\draw[line width=0.25mm,color=black] (A) to [bend left=-30] (B);
\coordinate (A) at (-0.6,0.5);
\coordinate (B) at (-0.3,0);
\draw[line width=0.25mm,color=black] (A) to [bend left=-30] (B);
\coordinate (A) at (0.6,0.5);
\coordinate (B) at (0.3,0);
\draw[line width=0.25mm,color=black] (B) to [bend left=-30] (A);
\node (2.9,0.6) at (1.1,0.2) {$=$};
\coordinate (A) at (2.4,0.5);
\coordinate (B) at (2.7,0);
\draw[line width=0.25mm,color=black] (A) to [bend left=-30] (B);
\coordinate (A) at (2.7,0);
\coordinate (B) at (3.3,0);
\draw[line width=0.25mm,color=black] (A) to [bend left=-30] (B);
\coordinate (A) at (3.6,0.5);
\coordinate (B) at (3.3,0);
\draw[line width=0.25mm,color=black] (B) to [bend left=-30] (A);
\node (2.9,0.6) at (1.8,0.25) {$\zeta^{-\alpha^{2}}$};
\node (2.9,0.6) at (3.35,0.25) {$\alpha$};
\node (0,0) at (3.9,0.2) {$.$};
\end{tikzpicture}\;
\end{center}
Taken together, the cap and cup correspond to the Dirac bra-ket.  This representation will be used in our two-string and four-string models of~\S\ref{sec:OtherModels}.

\subsection{Trace}\label{trace measure}
The normalized trace $\Tr (\cdot)$ is represented diagrammatically as
\begin{center}
\begin{tikzpicture}
\node (0,0) at (-1.75,0) {$\Tr$};
\node (0,0) at (-1.4,0) {$\bigg( $};
\draw (-1.1,-0.4) --(-1.1,0.5);
\node (0,0) at (-0.8,0) {$\bigg) $};
\node (0,0) at (-0.4,0) {$=$};
\node (0,0) at (0.1,0) {$\frac{1}{\delta}$};
\draw (0.8,0) circle [radius=0.5];
\node (0,0) at (1.65,0) {$=$};
\node (0,0) at (2,0) {{\large $1$}};
\node (0,0) at (2.2,-0.2) {$,$};
\end{tikzpicture} \; \; \; \; \; \; \; \; \; \; \; \; \; \; \; \; \; \,
\begin{tikzpicture}
\node (0,0) at (-1.85,0) {$\Tr$};
\node (0,0) at (-1.5,0) {$\bigg( $};
\node (0,0) at (-1.25,0.05) {k};
\draw (-1.1,-0.4) --(-1.1,0.5);
\node (0,0) at (-0.8,0) {$\bigg) $};
\node (0,0) at (-0.4,0) {$=$};
\node (0,0) at (0,0.05) {$\frac{1}{\delta}$};
\node (0,0) at (0.45,0.05) {k};
\draw (1.1,0) circle [radius=0.5];
\node (0,0) at (1.95,0) {$=$};
\node (0,0) at (2.3,0) {{\large $0$}};
\node (0,0) at (4.8,0) {$\mbox{for} \; \, 1 \leq k \leq d-1$};
\node (0,0) at (6.5,-0.15) {$.$};
\end{tikzpicture}
\end{center}
Here $\delta=\sqrt{d}$ represents the circle diagram constant,
\begin{center} \begin{tikzpicture}
\node (0,0) at (0,0.05) {$\delta$};
\node (0,0) at (0.5,0) {$=$};
\draw (1.4,0) circle [radius=0.5];
\node (0,0) at (2.1,-0.2) {$.$};
\end{tikzpicture} \end{center}

\subsection{Inner Product}
The standard, or computational, basis of the $\mathbb{Z}_{d}$ graded Hilbert space $\mathcal{H}_{d}(m)$  is $|i_{1}i_{2} \cdots i_{m} \rangle \equiv |i_{1} \rangle \otimes |i_{2}\rangle \otimes \cdots \otimes |i_{m} \rangle$,
for $0\leqslant i_{1},i_{2}, \ldots, i_{m}\leqslant d-1$.  This vector is graded by $\sum_{k=1}^{m} i_{k} \; \, \mbox{mod} \; \mbox{d}$.  
In our one-string model in \S\ref{sec:OtherModels}, we represent the vector $|i_{1}i_{2} \cdots i_{m} \rangle$ by
\[
\begin{tikzpicture}
\node (0,0) at (2,-0.4) {$i_{1}$};
\draw (2.2,-0.7) --(2.2,0.7);
\node (0,0) at (2.5,-0.2) {$i_{2}$};
\draw (2.7,-0.7) --(2.7,0.7);
\node (0,0) at (3.2,0) {$\cdots$};
\node (0,0) at (3.8,0.4) {$i_{m}$};
\draw (4,-0.7) --(4,0.7);
\node (0,0) at (4.5,-0.2) {$.$};
\end{tikzpicture}
\]
For  $a,b \in \mathcal{H}_{d}(m)$, we represent the inner product $\langle a|b \rangle$ by 
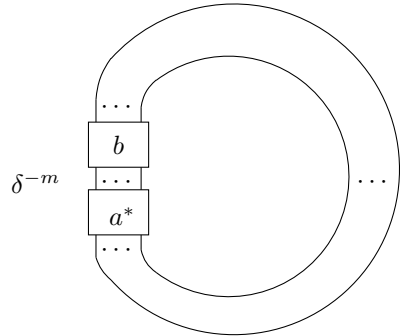
\begin{center}
\[
\begin{tikzpicture}
\node (0,0.65) at (1,0.7) {$\delta^{-m}$};
\draw (1.8,-0.3) --(1.8,0);
\coordinate (A) at (1.8,-0.3);
\coordinate (B) at (2,-0.6);
\draw[color=black] (A) to [bend left=-15] (B);
\draw (2.4,-0.2) --(2.4,0);
\coordinate (A) at (2.4,-0.2);
\coordinate (B) at (2.6,-0.5);
\draw[color=black] (A) to [bend left=-20] (B);
\node (0,0.65) at (2.1,-0.2) {$\cdots$};
\draw (1.7,0) rectangle +(0.8,0.6);
\node (0,0.65) at (2.15,0.25) {$a^{\ast}$};
\draw (1.8,0.6) --(1.8,0.9);
\draw (2.4,0.6) --(2.4,0.9);
\node (0,0.65) at (2.1,0.7) {$\cdots$};
\draw (1.7,0.9) rectangle +(0.8,0.6);
\node (0,0.65) at (2.1,1.2) {$b$};
\draw (1.8,1.5) --(1.8,1.8);
\coordinate (A) at (1.8,1.8);
\coordinate (B) at (2,2.35);
\draw[color=black] (A) to [bend left=15] (B);
\draw (2.4,1.5) --(2.4,1.7);
\coordinate (A) at (2.4,1.7);
\coordinate (B) at (2.6,2.06);
\draw[color=black] (A) to [bend left=20] (B);
\node (0,0.65) at (2.1,1.7) {$\cdots$};
\node (0,0.65) at (5.5,0.7) {$\cdots$};
\draw (2.6,-0.5) arc (-127:127:1.6);
\draw (2,-0.6) arc (-138:138:2.2);
\node (0,0.65) at (6.4,0.1) {$.$};
\end{tikzpicture} 
\]
\end{center}

\subsection{Partial Trace}
Planar parafermion algebras are half-braided, allowing a partial trace to be defined. The partial trace $\Tr^{j_{1},j_{2},\cdots,j_{k}}(\cdot)$ for $1 \leq j_{1},j_{2},\cdots,j_{k}\leq m$ is represented diagrammatically as
\begin{center}
\scalebox{0.85}{
\begin{tikzpicture}
\node (0,0) at (-4.75,0) {$\Tr^{j_{1},j_{2},\cdots,j_{k}}$};
\node (0,0) at (-3.75,0) {$\bigg( $};
\draw (-3.5,0.4) --(-2.1,0.4);
\draw (-3.5,-0.4) --(-3.5,0.4);
\draw (-3.5,-0.4) --(-2.1,-0.4);
\draw (-3.4,0.4) --(-3.4,0.7);
\draw (-3.2,0.4) --(-3.2,0.7);
\draw (-3.4,-0.4) --(-3.4,-0.7);
\draw (-3.2,-0.4) --(-3.2,-0.7);
\node (0,0) at (-2.8,0.55) {$\cdots$};
\node (0,0) at (-2.8,-0.55) {$\cdots$};
\node (0,0) at (-2.8,0) {$A$};
\draw (-2.4,0.4) --(-2.4,0.7);
\draw (-2.2,0.4) --(-2.2,0.7);
\draw (-2.4,-0.4) --(-2.4,-0.7);
\draw (-2.2,-0.4) --(-2.2,-0.7);
\draw (-2.1,-0.4) --(-2.1,0.4);
\node (0,0) at (-1.85,0) {$\bigg) $};
\node (0,0) at (-1.4,0) {$=$};
\node (0,0) at (-0.8,0) {$\frac{1}{{\delta}^{k}}$};
\draw (-0.2,-0.4) --(-0.2,0.4);
\draw (1.6,-0.4) --(1.6,0.4);
\draw (-0.2,0.4) --(1.6,0.4);
\draw (-0.2,-0.4) --(1.6,-0.4);
\draw (0,0.4) --(0,2.4);
\draw (0,-0.4) --(0,-2.2);
\draw (0.4,0.4) --(0.4,1);
\draw (0.4,1.3) --(0.4,2.4);
\draw (0.4,-0.4) --(0.4,-0.8);
\draw (0.4,-1) --(0.4,-2.2);
\draw (0.8,0.4) --(0.8,0.75);
\draw (0.8,0.95) --(0.8,1.5);
\draw (0.8,1.75) --(0.8,2.4);
\draw (0.8,-0.4) --(0.8,-0.7);
\draw (0.8,-0.9) --(0.8,-1.25);
\draw (0.8,-1.4) --(0.8,-2.2);
\draw (1.4,0.4) --(1.4,0.65);
\draw (1.4,0.85) --(1.4,1.3);
\draw (1.4,1.5) --(1.4,1.95);
\draw (1.4,2.1) --(1.4,2.4);
\draw (1.4,-0.4) --(1.4,-0.85);
\draw (1.4,-1) --(1.4,-1.25);
\draw (1.4,-1.4) --(1.4,-1.65);
\draw (1.4,-1.75) --(1.4,-2.2);
\draw (0.2,-0.4) arc (-164:173:2);
\draw (0.6,-0.4) arc (-164:165:1.5);
\draw (1,-0.4) arc (-164:148:1);
\node (0,0) at (0.70,-0.15) {$A$};
\node (0,0) at (0.18,0.18) {$ \scriptstyle j_{1}$};
\node (0,0) at (0.58,0.18) {$ \scriptstyle j_{2}$};
\node (0,0) at (0.87,0.18) {$\scriptstyle \cdots$};
\node (0,0) at (1.2,0.18) {$\scriptstyle  j_{k}$};
\node (0,0) at (4.4,-0.3) {$.$};
\end{tikzpicture}}
\end{center}
\noindent On the right hand side the $j_{1},j_{2},\cdots,j_{k}$ strings are closed to form caps. The nonclosed strings always move under the caps. Moreover, the strings are  closed clockwise from top to bottom. The spherical condition allows strings to be closed counterclockwise from top to bottom. See \S $2.2$ of \cite{JL} for details and the definition of the spherical condition.

\subsection{Measurement} \label{measure}
\begin{figure}[h]
\centering
\begin{tikzpicture}
\node (5.3,0.3) at (0.15,0.8) {$\scriptstyle j_{1}$};
\draw (0.15,1) --(0.15,2);
\draw (-0.15,1.15) --(-0.15,2);
\draw (-0.15,-0.15) --(-0.15,-1);
\node (5.3,0.3) at (0.75,0.8) {$\scriptstyle j_{2}$};
\draw (0.75,1) --(0.75,2);
\draw (0.45,1.15) --(0.45,2);
\draw (0.45,-0.15) --(0.45,-1);
\node (5.3,0.3) at (1.1,0.8) {$\scriptstyle \cdots$};
\node (5.3,0.3) at (1.45,0.8) {$\scriptstyle j_{k}$};
\draw (1.35,1) --(1.35,2);
\draw (1.05,1.15) --(1.05,2);
\draw (1.05,-0.15) --(1.05,-1);
\draw (1.65,1.15) --(1.65,2);
\draw (1.65,-0.15) --(1.65,-1);
\draw (-0.3,0) --(1.8,0);
\draw (-0.3,0) --(-0.3,1);
\draw (1.8,0) --(1.8,1);
\draw (-0.3,1) --(1.8,1);
\draw (1.15,0.1) arc (0:180:.4);
\draw (0.7,0.1) --(1,0.6);
\node (5.3,0.3) at (2.4,0) {$$};
\end{tikzpicture} 
\caption{Measurement: unmeasured strings pass underneath the meter.
\label{meter string}
}
\centering
\end{figure}
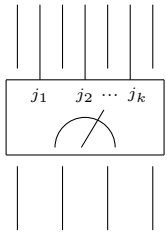

We use the meter in Figure \ref{meter}
to perform a measurement of the strings  $j_{1},j_{2},\cdots,j_{k}$, represented diagrammatically in Figure \ref{j strings}.
The result of the measurement is represented  in Figure \ref{meter string}.
\begin{figure}[h]
\begin{center}
\begin{tikzpicture}
\draw (0,0) --(1.5,0);
\draw (0,0) --(0,1);
\draw (1.5,0) --(1.5,1);
\draw (0,1) --(1.5,1);
\draw (1.15,0.3) arc (0:180:.4);
\draw (0.7,0.3) --(1,0.8);
\end{tikzpicture}
\caption{Meter. \label{meter}}
\end{center}
\end{figure}
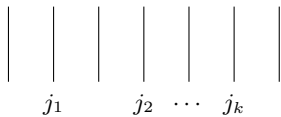
\begin{figure}[h]
\begin{center}
\begin{tikzpicture}
\node (5.3,0.3) at (0,0.7) {$j_{1}$};
\draw (-0.6,1) --(-0.6,2);
\draw (0,1) --(0,2);
\node (5.3,0.3) at (1.2,0.7) {$j_{2}$};
\draw (0.6,1) --(0.6,2);
\draw (1.2,1) --(1.2,2);
\draw (1.8,1) --(1.8,2);
\node (5.3,0.3) at (1.8,0.7) {$\cdots$};
\node (5.3,0.3) at (2.4,0.7) {$j_{k}$};
\draw (2.4,1) --(2.4,2);
\draw (3,1) --(3,2);
\node (5.3,0.3) at (3.5,1.4) {$$};
\end{tikzpicture}
\caption{Strings to be measured, $j_{1},j_{2},\cdots,j_{k}$ \label{j strings}.}
\end{center}
\end{figure}

\noindent The meter designates that the $j_{1},j_{2},\cdots,j_{k}$ strings are to be closed from top to bottom to form caps. We proceed by removing the meter from the diagram and closing the $j_{1},j_{2},\cdots,j_{k}$ strings, as discussed above for the partial trace.  

Let us illustrate a measurement for parafermions of order $d$. Consider three parafermions in the computational basis
\begin{center} \begin{tikzpicture}
\node (0,0) at (2,-0.4) {$i_{1}$};
\draw (2.2,-0.7) --(2.2,0.7);
\node (0,0) at (2.5,0) {$i_{2}$};
\draw (2.7,-0.7) --(2.7,0.7);
\node (0,0) at (3,0.4) {$i_{3}$};
\draw (3.2,-0.7) --(3.2,0.7);
\node (0,0) at (3.5,-0.3) {$.$};
\end{tikzpicture} \end{center}
Suppose we want to measure the first two tensor factors. We place the meter under the first two strings as illustrated below,
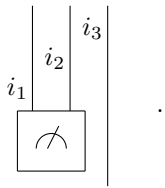
\begin{center} \begin{tikzpicture}
\node (0,0) at (2,-0.4) {$i_{1}$};
\draw (2.2,-0.7) --(2.2,0.7);
\node (0,0) at (2.5,0) {$i_{2}$};
\draw (2.7,-0.7) --(2.7,0.7);
\draw (2,-0.7) --(2.9,-0.7);
\draw (2,-0.7) --(2,-1.5);
\draw (2.9,-0.7) --(2.9,-1.5);
\draw (2,-1.5) --(2.9,-1.5);
\draw (2.65,-1.2) arc (0:180:.2);
\draw (2.4,-1.2) --(2.55,-0.9);
\node (0,0) at (3,0.4) {$i_{3}$};
\draw (3.2,-1.7) --(3.2,0.7);
\node (0,0) at (3.9,-0.7) {$.$};
\end{tikzpicture} \end{center}
\noindent The meter under the first two strings designates that those strings are closed to form caps. We obtain
\begin{center} \begin{tikzpicture}
\node (0,0) at (0,0) {$\frac{1}{\delta^{2}}$};
\node (0,0) at (0.55,0.05) {$i_{1}$};
\draw (1.2,0) circle [radius=0.5];
\node (0,0) at (2,0.2) {$i_{2}$};
\draw (2.65,0) circle [radius=0.5];
\node (0,0) at (3.4,0.4) {$i_{3}$};
\draw (3.6,-0.7) --(3.6,0.7);
\node (0,0) at (4.1,-0.3) {$.$};
\end{tikzpicture} \end{center}
As the circle diagram has value $\delta$, the measurement of an unconnected string with no excitation is normalized to give the value $1$.

\section{Braids and Entanglement}\label{Particles}
\subsection{Background}
The topological approach to quantum computation became important with Kitaev's 1997 paper proposing an anyon computer---work that only appeared some five years later in print~\cite{anyon computer}. In \S6 on the arXiv, he described the braiding and fusing of anyonic excitations in a fault-tolerant way. Freedman, Kitaev, Larsen, and Wang explored braiding further~\cite{topological quantum}, motivated by the pioneering work of Jones, Atiyah, and Witten on knots and topological field theory~\cite{Jon87, Ati88, Wit88}.\footnote{Diagrammatic notation  in quantum information theory originated in the quantum circuit model of Deutsch \cite{D}, although without the consideration of topology.} 
Kauffman and Lomonaco remarked that the braid diagram describes maximal entanglement \cite{KL-2}.

\subsection{The Braid}  
For fermions and parafermions the parafermionic Fock space $\mathcal{H}_{d} (m)$ is isomorphic to the $m$-qudit  space $\mathbb{C}^{d^{m}}$, where $d$ denotes the parafermion order and $m$ is the number of modes. The choice of $d=2$ is the standard Fock space for $m$ fermionic modes, which is isomorphic to the $m$-qubit space \cite{jordan wigner}.

The notion of fermionic entanglement for pure states was analyzed in \cite{DoubleOccupancy,quantum correlations fermions,fermion entanglement}, whereby product states are those that one can write as a tensor product in the Fock representation. This definition of entanglement naturally generalizes to the case of parafermionic pure states. We refer to this generalization as \textit{parafermionic entanglement}.

The unitary braid operator\footnote{These braids can be ``Baxterized'' in the sense of Jones \cite{Jon91}.  They are the limits of solutions to the Yang-Baxter equation in statistical physics \cite{Yang,Baxter}, and have actually been introduced earlier in \cite{F-Z}.  
%However related topological aspects of braiding in statistical physics and in mathematics had been investigated many years earlier.  
Such kinds of braid statistics in field theory and quantum Hall systems were considered extensively by Fr\"ohlich, see \cite{Froehlich,FroehlichCargese}.} in Figure \ref{Braid-Entanglement} canonically generates maximal fermionic and parafermionic entanglement for arbitrary finite dimensions. See \S $8$ of \cite{JL} for details and the definition of the braid

\begin{figure}[h]
\centering \begin{tikzpicture}
\draw (1.25,0.5) --(1,0.75);
\draw (1.75,0) --(1.5,0.25);
\draw (1,0) --(1.75,0.75);
\node (2.45,0.3) at (2.2,0.3) {$\equiv$};
\node (3.4,0.3) at (3.4,0.4) {$\frac{\omega^{-\frac{1}{2}}}{\sqrt{d}}  \, 
 \sum_{k=0}^{d-1}$};
\node (0,0) at (4.6,0.7) {k};
\draw (4.75,-0.1) --(4.75,0.9);
\node (0,0) at (5.3,0.2) {$-k$};
\draw (5.6,-0.1) --(5.6,0.9);
\end{tikzpicture} \centering
\caption{Braid diagram for entanglement.
\label{Braid-Entanglement}
}
\end{figure}

\noindent where $\omega = \frac{1}{\sqrt{d}} \sum_{j=0}^{d-1} \zeta^{j^{2}}$ is a phase. (Recall  $\zeta^{2}=q$, and $q^{d}=\zeta^{d^{2}}=1$.)  The braid has the special property that qudit excitations can move under the braid crossing as illustrated in Figure \ref{Braid-qudit}.

Since the braid is unitary,  its adjoint equals the inverse braid
\begin{center}
\begin{equation}
\begin{tikzpicture}
\draw (1.5,6.5) --(1.75,6.75);
\draw (1,6.75) --(1.75,6);
\draw (1,6) --(1.25,6.25);
\node (2.45,0.3) at (2.2,6.3) {$\equiv$};
\node (3.4,0.3) at (3.4,6.4) {$\frac{\omega^{\frac{1}{2}}}{\sqrt{d}}  \, \sum_{k=0}^{d-1}$};
\node (0,0) at (4.6,6.2) {k};
\draw (4.75,5.9) --(4.75,6.9);
\node (0,0) at (5.3,6.7) {$-k$};
\draw (5.6,5.9) --(5.6,6.9);
\node (5.3,0.3) at (5.9,6.2) {$.$};
\end{tikzpicture} 
\end{equation}
\end{center}
This inverse disentangles fermionic and parafermionic states of arbitrary finite dimension in a canonical way.

In the following example we illustrate maximal entanglement for the fermionic case $d=2$. Consider the $2$-qubit space
\begin{center} \begin{tikzpicture}
\node (0,0) at (-2.6,0.7) {$\mathbb{C}^{4} = \mbox{Span}_{\mathbb{C}} \bigg\{$};
\draw (-1.3,0.3) --(-1.3,1.1);
\draw (-1.1,0.3) --(-1.1,1.1);
\node (0,0) at (-0.9,0.3) {,};
\draw (-0.7,0.3) --(-0.7,1.1);
\node (0,0) at (-0.45,0.7) {$1$};
\draw (-0.35,0.3) --(-0.35,1.1);
\node (0,0) at (-0.15,0.3) {,};
\node (0,0) at (0.1,0.7) {$1$};
\draw (0.2,0.3) --(0.2,1.1);
\draw (0.4,0.3) --(0.4,1.1);
\node (0,0) at (0.6,0.3) {,};
\node (0,0) at (0.9,0.55) {$1$};
\draw (1,0.3) --(1,1.1);
\node (0,0) at (1.4,0.9) {$-1$};
\draw (1.65,0.3) --(1.65,1.1);
\node (0,0) at (1.95,0.7) {$\bigg\}$};
\node (0,0) at (2.1,0.2) {,};
\end{tikzpicture} \end{center}
 \noindent in which the braid acts on the basis by
\begin{equation}\label{Braid-1}
\; \; \; \; \;
\begin{tikzpicture}
\draw (1.25,0.5) --(1,0.75);
\draw (1.75,0) --(1.5,0.25);
\draw (1,0) --(1.75,0.75);
\draw (1,0.75) --(1,1.5);
\draw (1.75,0.75) --(1.75,1.5);
\node (2,0.75) at (2.25,0.6) {$=$};
\node (2.9,0.6) at (2.9,0.65) {$\frac{\omega^{{-\frac{1}{2}}}}{\sqrt{2}}$};
\node (2.9,0.6) at (3.4,0.6) {$\bigg( $};
\draw (3.6,0.3) --(3.6,1);
\draw (3.8,0.3) --(3.8,1);
\node (3.8,0.3) at (4.2,0.6) {$-$};
\node (0,0) at (4.6,0.5) {$1$};
\draw (4.75,0.3) --(4.75,1);
\node (0,0) at (5.1,0.75) {$-1$};
\draw (5.4,0.3) --(5.4,1);
\node (5.5,0.6) at (5.6,0.6) {$\bigg) $};
\node (0,0) at (6,0.4) {,};
\end{tikzpicture}  
\end{equation}
\begin{equation}\label{Braid-2}
 \; \; \;
\begin{tikzpicture}
\draw (1.25,0.5) --(1,0.75);
\draw (1.75,0) --(1.5,0.25);
\draw (1,0) --(1.75,0.75);
\draw (1,0.75) --(1,1.5);
\node (0,0) at (1.65,1.2) {$1$};
\draw (1.75,0.75) --(1.75,1.5);
\node (2,0.75) at (2.25,0.6) {$=$};
\node (2.9,0.6) at (2.9,0.65) {$\frac{\omega^{{-\frac{1}{2}}}}{\sqrt{2}}$};
\node (2.9,0.6) at (3.4,0.6) {$\bigg( $};
\draw (3.6,0.3) --(3.6,1);
\node (0,0) at (3.8,0.6) {$1$};
\draw (3.95,0.3) --(3.95,1);
\node (3.8,0.3) at (4.35,0.6) {$-$};
\node (0,0) at (4.75,0.6) {$1$};
\draw (4.9,0.3) --(4.9,1);
\draw (5.1,0.3) --(5.1,1);
\node (5,0.6) at (5.3,0.6) {$\bigg) $};
\node (0,0) at (5.7,0.4) {,}; \end{tikzpicture} 
\end{equation}
\begin{equation} \label{Braid-3}
\begin{tikzpicture}
\draw (1.25,0.5) --(1,0.75);
\draw (1.75,0) --(1.5,0.25);
\draw (1,0) --(1.75,0.75);
\node (0,0) at (0.85,1.2) {$1$};
\draw (1,0.75) --(1,1.5);
\draw (1.75,0.75) --(1.75,1.5);
\node (2,0.75) at (2.25,0.6) {$=$};
\node (2.9,0.6) at (2.9,0.65) {$\frac{\omega^{{-\frac{1}{2}}}}{\sqrt{2}}$};
\node (2.9,0.6) at (3.4,0.6) {$\bigg( $};
\node (0,0) at (4.75,0.6) {$1$};
\draw (4.9,0.3) --(4.9,1);
\draw (5.1,0.3) --(5.1,1);
\node (3.8,0.3) at (4.35,0.6) {$+$};
\draw (3.6,0.3) --(3.6,1);
\node (0,0) at (3.8,0.6) {$1$};
\draw (3.95,0.3) --(3.95,1);
\node (4.95,0.6) at (5.3,0.6) {$\bigg) $};
\node (0,0) at (5.7,0.4) {,}; \end{tikzpicture} \end{equation}
\begin{equation}\label{Braid-4}
\; \; \; \;\begin{tikzpicture}
\draw (1.25,0.5) --(1,0.75);
\draw (1.75,0) --(1.5,0.25);
\draw (1,0) --(1.75,0.75);
\node (0,0) at (0.85,1) {$1$};
\draw (1,0.75) --(1,1.5);
\node (0,0) at (1.5,1.25) {$-1$};
\draw (1.75,0.75) --(1.75,1.5);
\node (2,0.75) at (2.25,0.6) {$=$};
\node (2.9,0.6) at (2.9,0.65) {$\frac{\omega^{{-\frac{1}{2}}}}{\sqrt{2}}$};
\node (2.9,0.6) at (3.4,0.6) {$\bigg( $};
\node (0,0) at (4.6,0.5) {$1$};
\draw (4.75,0.3) --(4.75,1);
\node (0,0) at (5.1,0.75) {$-1$};
\draw (5.4,0.3) --(5.4,1);
\node (3.8,0.3) at (4.2,0.6) {$+$};
\draw (3.6,0.3) --(3.6,1);
\draw (3.8,0.3) --(3.8,1);
\node (5.4,0.6) at (5.6,0.6) {$\bigg) $};
\node (0,0) at (6,0.4) {.}; \end{tikzpicture}
\end{equation}

In quantum computation the braid is ``imprimitive'' in the sense of the Brylinskis, since it is \textit{entangling}.
This result yields a partial topological quantum computer for fermions and parafermions. Additionally, the braid may be applied to  construct several quantum information protocols diagrammatically, which  consume entanglement as a resource.

In Figure \ref{Braid-bi} we illustrate the braid $b_i$  on the $i$th and $i+1$th strings.
\begin{figure}[h]
\begin{tikzpicture}
\draw (1.25,2.5) --(1,2.75);
\draw (1.75,2) --(1.5,2.25);
\draw (1,2) --(1.75,2.75);
\node  (3.75,0.35) at (0.5,2.4) {$b_{i}=$};
\node  (3.75,0.35) at (0.95,1.75) {$i$};
\node  (3.75,0.35) at (1.8,1.75) {$i+1$};
\end{tikzpicture}
\caption{Braid between Adjacent Strings.
\label{Braid-bi}
}
\end{figure}

\section{Qudit Pauli $X,Y,Z$ Matrices}\label{XYZ Matrices}
One can find qudit, Pauli $X,Y,Z$ matrices that satisfy the relations 
	\be\label{PauliRelations1a}
	X^d=Y^d =Z^d=\,1\;,
	\ee
	\be	\label{PauliRelations1b}
	YX=q\,XY\;,\quad
	ZY=q\,YZ\;, \quad\text{and}\quad
	XZ=q\,ZX\;.
	\ee
Here $q=e^{\frac{2\pi i}{d}}$.  These matrices must also satisfy a second set of relations defined in terms of a square root $\zeta=q^{\frac{1}{2}}$ for which $\zeta^{d^{2}}=1$, namely
	\begin{equation}\label{PauliRelations2}
	 XYZ = YZX = ZXY =\zeta^{-1} \;.
	\end{equation}
	
In \S4 of \cite{JL} we give two different solutions $\widehat{X},\widehat{Y},\widehat{Z}$ for $X,Y,Z$. Each solution is a quadratic function of four qudit generators $c_{1},c_{2},c_{3},c_{4}$ of the parafermion algebra. 

\subsection{Solution I}  \label{Solution1}  In \S5 of \cite{JL}  we give a diagrammatic interpretation for these operators.  Our first solution has the form  
	\be\label{xyz1}
	\widehat{X} = \zeta\, c_{1}^{-1}c_{4} \;,\quad
	\widehat{Y} = \zeta\, c_{2}c_{4}^{-1} \;,\quad
	\widehat{Z} = \zeta\, c_{3}^{-1}c_{4} \;.
	\ee
These matrices satisfy relations \eqref{PauliRelations1a}--\eqref{PauliRelations1b}, but they do not identically satisfy  \eqref{PauliRelations2} on the entire Hilbert space.  

As explained in \cite{JL},  the product $\widehat{X}\widehat{Y}\widehat{Z}$ has the form
	\be\label{Gamma-Relation}
	\widehat{X}\widehat{Y}\widehat{Z} =
	\widehat{Y}\widehat{Z}\widehat{X} =
	\widehat{Z}\widehat{X}\widehat{Y} = \zeta^{-1} \gamma\;,
	\ee
where
	\be
	\gamma =q c_{1}^{-1}c_{2}c_{3}^{-1}c_{4}=e^{iQ}\;.
	\ee
This defines the self-adjoint charge operator $Q$ mod $\mathbb{Z}_{d}$.  Since $\widehat{X}$, $\widehat{Y}$, $\widehat{Z}$ are zero-graded, each operator acts on the eigenspaces of $\gamma$.   One achieves the missing relations  \eqref{PauliRelations2} by restricting to the charge-zero subspace for which $\gamma=+1$.

\subsection{Solution II} 
Our second solution is
	\be\label{xyz2}
	\widehat{X} = \zeta\, c_{1}^{-1}c_{2} \;,\quad
	\widehat{Y} = \zeta\, c_{1}c_{3}^{-1} \;,\quad
	\widehat{Z} = \zeta\, c_{1}^{-1}c_{4} \;.
	\ee
These matrices $\widehat{X}$, $\widehat{Y}$, $\widehat{Z}$ also satisfy the relations  \eqref{PauliRelations1a}--\eqref{PauliRelations1b} and \eqref{Gamma-Relation}. So they satisfy \eqref{PauliRelations2} on the same eigenspace $\gamma=+1$.
One can perform this construction at each one of various sites labelled by a subscript $j$, giving rise to a representation of operators $\widehat{X}_{j},\widehat{Y}_{j},\widehat{Z}_{j}$ at each site, and that mutually commute at different sites.

Diagrammatically our two solutions \eqref{xyz1} and \eqref{xyz2} lead to very different looking models, which in  \S\ref{sec:OtherModels} we call four-string models (each string representing a qudit) of type I and type II.  In the related paper \cite{JL}, we give details and develop these results in a more general context.

\subsection{Comparison with Kitaev's $d=2$ construction}
Solution~I, given in \S \ref{Solution1},  is related to the construction of Kitaev for $d=2$. Equation (11) of \cite{Kitaev-2} gives the $d=2$ representation that one commonly uses in condensed-matter physics, in which 
	\[
	\widehat{X} = i\, c_{1}c_{4} \;,\quad
	\widehat{Y} = i\, c_{2}c_{4} \;,\quad
	\widehat{Z} = i\, c_{3}c_{4} \;,
	\]
and one has  
	\[
	\widehat{X}\widehat{Y}\widehat{Z} =
	\widehat{Y}\widehat{Z}\widehat{X} =
	\widehat{Z}\widehat{X}\widehat{Y} = i c_{1}c_{2}c_{3}c_{4}\;.
	\]
One takes
	\[
	\widehat{X}\widehat{Y}\widehat{Z} =
	\widehat{Y}\widehat{Z}\widehat{X} =
	\widehat{Z}\widehat{X}\widehat{Y} = i \;,
	\]
on the subspace for which  $c_{1}c_{2}c_{3}c_{4}=1$.

We can recover this solution of Kitaev from our formulas, by taking $d=2$, $\zeta=i$, and $\gamma=-1$ (rather than $\gamma=+1$ as we require). 
Likewise one can generalize this construction for arbitrary $d$, by taking  
	\[
	\widehat{X}\widehat{Y}\widehat{Z} =
	\widehat{Y}\widehat{Z}\widehat{X} =
	\widehat{Z}\widehat{X}\widehat{Y} = \zeta^{-1}q^{k} \;,
	\]
on the subspace graded by $k$ mod $d$, where $\gamma=q^{k}$.  

However our Solution I is different in a subtle way from Kitaev's  construction.  
In our four-string model described in  \S\ref{Sect:DifferentModels-4}, we represent a qudit by a charge-zero, particle-anti-particle pair. The neutral total charge means that $\gamma=+1$. 

For this reason we find our choice natural.  
With our basis, the qudit Pauli $X,Y,Z$ are neutral and act as $d\times d$ matrices in a natural way, preserving charge neutrality.  
But in  Kitaev's model $\gamma=-1$, so qudits are not neutral. Then  one loses the particle-anti-particle interpretation of qudits, that we exploit in our diagrams.  

\section{Protocols}\label{Protocols}
\subsection{Entanglement Distribution Protocol} We apply the braid to construct the entanglement distribution protocol. Consider the computational basis for two parafermions of order $d$,
\begin{center}
\begin{tikzpicture}
\node (0,0) at (2,-0.2) {$i_{1}$};
\draw (2.2,-0.7) --(2.2,0.7);
\node (0,0) at (2.5,0.2) {$i_{2}$};
\draw (2.7,-0.7) --(2.7,0.7);
\node (0,0) at (6.2,0) {$\mbox{for} \; \; \; \; 0 \leq i_{1},i_{2}\leq d-1 $};
\node (0,0) at (8.3,-0.15) {$.$};
\end{tikzpicture}
\end{center}
\noindent We act with the braid of Figure \ref{Braid-Entanglement} to generate maximal entanglement, namely

%\begin{centering}
\begin{equation} \begin{tikzpicture}
\node (0,0) at (0.8,1.1) {$i_{1}$};
\draw (1,1.75) --(1,0.75);
\node (0,0) at (1.55,1.4) {$i_{2}$};
\draw (1.75,1.75) --(1.75,0.75);
\draw (1.25,0.5) --(1,0.75);
\draw (1.75,0) --(1.5,0.25);
\draw (1,0) --(1.75,0.75);
\node (2.45,0.3) at (2.3,0.9) {$=$};
\node (3.4,0.3) at (3.1,1) {$\frac{\omega^{-\frac{1}{2}}}{\sqrt{d}}$};
\node (3.4,0.3) at (4.5,1) {${\displaystyle \sum_{k=0}^{d-1} q^{(k+i_{1})k}}$};
\node (0,0.2) at (6,0.7) {$k+i_{1}$};
\draw (6.55,0.2) --(6.55,1.8);
\node (0,0) at (7.6,1.3) {$-k+i_{2}$};
\draw (8.4,0.2) --(8.4,1.8);
\node (0,0.2) at (8.8,0.6) {$.$};
\end{tikzpicture} 
\label{Channel}
\end{equation}
%\end{centering}

\noindent The special case of fermions was shown in \eqref{Braid-1}--\eqref{Braid-4}. The remaining step of the protocol involves distribution of the entanglement through a noiseless quantum channel \cite{quantum information theory}. Such a channel leaves \eqref{Channel} invariant. Physically the distribution is performed by a variety of methods \cite{free space, entanglement distribution}.

\subsection{Entanglement-Swapping Protocol\label{sec:ESwap}} We can also apply the braid to construct the entanglement-swapping protocol. This protocol inputs four disentangled fermionic or parafermionic states, and maximally entangles two of the states without trivially braiding them. Physically these entangled states do not need to share any common past \cite{bell experiment,DeterministicEntanglementSwapping}.

Consider the diagram below, which entangles the first and second strings, and it entangles the third and fourth strings. Then, it disentangles the second and third strings with the inverse braid:
\begin{center}
\begin{equation}
\begin{tikzpicture}
\draw (1.25,2.5) --(1,2.75);
\draw (2.25,1.5) --(1.5,2.25);
\draw (1,2) --(1.75,2.75);
\draw (2.25,2.5) --(2,2.75);
\draw (2.75,2) --(2.5,2.25);
\draw (2,2) --(2.75,2.75);
\draw (1.5,1.5) --(1.75,1.75);
\draw (1,1.5) --(1,2);
\draw (2.75,1.5) --(2.75,2);
\node (5.3,0.3) at (3.3,1.8) {$.$};
\end{tikzpicture} 
\label{TangleUntangle}
\end{equation}
\end{center}
\noindent We proceed by placing the meter, introduced in \S \ref{measure}, under the second and third strings of \eqref{TangleUntangle}, as illustrated below
\begin{center}
\begin{equation}
\begin{tikzpicture}
\draw (1.25,2.5) --(1,2.75);
\draw (2.25,1.5) --(1.5,2.25);
\draw (1,2) --(1.75,2.75);
\draw (2.25,2.5) --(2,2.75);
\draw (2.75,2) --(2.5,2.25);
\draw (2,2) --(2.75,2.75);
\draw (1.5,1.5) --(1.75,1.75);
\draw (1,0.8) --(1,2);
\draw (2.75,0.8) --(2.75,2);
\draw (2.5,1.5) --(1.25,1.5);
\draw (1.25,1.5) --(1.25,0.8);
\draw (2.5,1.5) --(2.5,0.8);
\draw (2.5,0.8) --(1.25,0.8);
\draw (2.15,1) arc (0:180:.25);
\draw (1.85,1) --(2.1,1.4);
\node (5.3,0.3) at (3.15,1.5) {$.$};
\end{tikzpicture} 
\label{Measurement}
\end{equation}
\end{center}
\noindent We claim that \eqref{Measurement} acts by maximally entangling the leftmost and rightmost input states as desired. Here we use the relations in \S8 of~\cite{JL}.

\begin{figure}[h]
\begin{center}
\begin{tikzpicture}
\draw (1.25,2.5) --(1,2.75);
\draw (2.25,1.5) --(1.5,2.25);
\draw (1,2) --(1.75,2.75);
\draw (2.25,2.5) --(2,2.75);
\draw (2.75,2) --(2.5,2.25);
\draw (2,2) --(2.75,2.75);
\draw (1.5,1.5) --(1.75,1.75);
\draw (1,0.8) --(1,2);
\draw (2.75,0.8) --(2.75,2);
\draw (2.5,1.5) --(1.25,1.5);
\draw (1.25,1.5) --(1.25,0.8);
\draw (2.5,1.5) --(2.5,0.8);
\draw (2.5,0.8) --(1.25,0.8);
\draw (2.15,1) arc (0:180:.25);
\draw (1.85,1) --(2.1,1.4);
\draw[->] (3.6,1.8) --(4.2,1.8);
\draw (5.35,1.9) --(5,2.25);
\draw (6,1.25) --(5.65,1.6);
\draw (5,1.25) --(6,2.25);
\end{tikzpicture}
\caption{Entanglement-swapping diagram. \label{Measurementbraid}}
\end{center}
\end{figure}
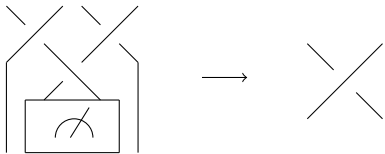

We remove the meter in \eqref{Measurement}, closing the second and third strings to form caps as illustrated below
\begin{center}
\begin{equation}
\begin{tikzpicture}
\node (0,0) at (4.4,1.6) {$\frac{1}{{\delta}^{2}}$};
\draw (5.25,2.5) --(5,2.75);
\draw (5.75,2) --(5.5,2.25);
\draw[color=red]  (5.1,2.2) --(5.75,2.75);
\coordinate (A) at (5.1,2.2);
\coordinate (B) at (5,2);
\draw[color=red] (A) to [bend left=-20] (B);
\draw (7.65,2.2) --(7.5,2.3);
\coordinate (A) at (7.65,2.2);
\coordinate (B) at (7.75,2);
\draw[color=black] (A) to [bend left=20] (B);
\draw[color=red]  (7,2) --(7.75,2.75);
\draw (7.25,2.5) --(7,2.75);
\draw[color=red]  (5,0.8) --(5,2);
\draw (7.75,0.8) --(7.75,2);
\coordinate (A) at (5.75,2.75);
\coordinate (B) at (6.15,2.6);
\draw[color=red] (A) to [bend left=60] (B);
\draw[color=red]  (6.15,1) --(6.15,2.6);
\coordinate (A) at (5.78,0.8);
\coordinate (B) at (6.15,1);
\draw[color=red] (A) to [bend left=-60] (B);
\coordinate (A) at (5.78,0.8);
\coordinate (B) at (5.82,1);
\draw[color=red] (A) to [bend left=50] (B);
\coordinate (A) at (6.6,2.6);
\coordinate (B) at (7,2.75);
\draw[color=black] (A) to [bend left=60] (B);
\draw (6.6,1) --(6.6,2.6);
\coordinate (A) at (6.6,1);
\coordinate (B) at (7,0.8);
\draw[color=black] (A) to [bend left=-60] (B);
\coordinate (A) at (7,0.8);
\coordinate (B) at (7.04,1);
\draw[color=black] (A) to [bend left=-60] (B);
\draw (6.1,1.6) --(5.75,2);
\draw (6.55,1.2) --(6.2,1.5);
\draw (7.04,1) --(6.65,1.17);
\draw[color=red]  (6.65,1.63) --(7,2);
\draw[color=red]  (6.4,1.4) --(6.55,1.55);
\draw[color=red]  (6.2,1.22) --(6.33,1.33);
\draw[color=red]  (5.82,1) --(6.1,1.18);
\node (5.3,0.3) at (8.2,1.5) {$.$};
\end{tikzpicture}
\label{Swapping}
\end{equation} \end{center}
\noindent Isotopy is a property of parafermion planar algebra. This topological notion simplifies the computation of \eqref{Swapping} and reduces it to a scalar multiple of the braid. It permits us to move the strings in three-dimensional space. We note that the red string under the Reidemeister moves becomes the braid's over crossing. We use the second Reidemeister move on \eqref{Swapping} to obtain
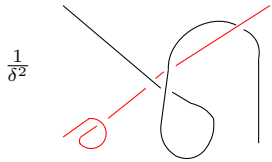
\begin{center}
\begin{equation*}
\begin{tikzpicture}
\node (0,0) at (4.4,-0.8) {$\frac{1}{{\delta}^{2}}$};
\coordinate (A) at (7.6,-0.7);
\coordinate (B) at (7.4,-0.32);
\draw[color=black] (A) to [bend left=-40] (B);
\coordinate (C) at (7.3,-0.26);
\coordinate (D) at (6.4,-0.85);
\draw[color=black] (C) to [bend left=-60] (D);
\draw (6.3,-1.6) --(6.4,-0.85);
\coordinate (E) at (6.3,-1.6);
\coordinate (F) at (6.5,-2);
\draw[color=black] (E) to [bend left=-30] (F);
\coordinate (G) at (6.5,-2);
\coordinate (H) at (6.9,-1.9);
\draw[color=black] (G) to [bend left=-40] (H);
\coordinate (I) at (6.9,-1.9);
\coordinate (J) at (7,-1.5);
\draw[color=black] (I) to [bend left=-20] (J);
\coordinate (K) at (7,-1.5);
\coordinate (L) at (6.7,-1.3);
\draw[color=black] (K) to [bend left=-20] (L);
\draw (6.7,-1.3) --(6.45,-1.15);
\draw (6.3,-1.1) --(5,-0);
\draw (7.6,-0.7) --(7.6,-1.83);
\draw[color=red] (5,-1.75) --(5.35,-1.5);
\draw[color=red] (5.22,-1.83) arc (-140:100:.2);
\draw[color=red] (5.233,-1.75) arc (-212:-150:.095);
\draw[color=red] (5.233,-1.75) --(5.45,-1.6);
\draw[color=red] (5.6,-1.5) --(6.1,-1.1);
\draw[color=red] (6.23,-0.97) --(6.34,-0.88);
\draw[color=red] (6.5,-0.8) --(7.75,0);
\node (5.3,0.3) at (8.4,-1) {$.$};
\end{tikzpicture}
\end{equation*}
\end{center}
\noindent Application of the second and third Reidemeister moves simplifies the diagram above to
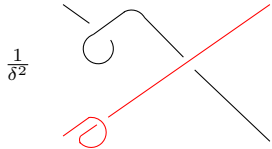
\begin{center}
\begin{equation*}
\begin{tikzpicture}
\node (0,0) at (4.4,-0.8) {$\frac{1}{{\delta}^{2}}$};
\draw (5.35,-0.25) --(5,0);
\draw (5.4,-0.4) arc (-250:35:.2);
\draw (5.4,-0.4) --(5.82,-0.095);
\draw (5.81,-0.1) arc (120:30:.2);
\draw (6.08,-0.17) --(6.6,-0.7);
\draw (6.75,-0.87) --(7.75,-1.82);
\draw[color=red] (5,-1.75) --(5.35,-1.5);
\draw[color=red] (5.22,-1.83) arc (-140:100:.2);
\draw[color=red] (5.233,-1.75) arc (-212:-150:.095);
\draw[color=red] (5.233,-1.75) --(5.45,-1.6);
\draw[color=red] (5.6,-1.5) --(7.75,0);
\node (5.3,0.3) at (8.4,-1) {$.$};
\end{tikzpicture}
\end{equation*}
\end{center}
The braid and its inverse in the last diagram above have opposite coefficients by the first Reidemeister move, reducing the diagram to
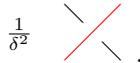
\begin{center}
\begin{equation*}
\begin{tikzpicture}
\node (0,0) at (4.4,-3.4) {$\frac{1}{{\delta}^{2}}$};
\draw[color=red] (5.75,-3) --(5,-3.75);
\draw (5,-3) --(5.25,-3.25);
\draw (5.5,-3.5) --(5.75,-3.75);
\node (0,0) at (6,-3.75) {$.$};
\end{tikzpicture}
\end{equation*}
\end{center}
\noindent Therefore, the entanglement-swapping diagram in \eqref{Measurement} maximally entangles the leftmost and rightmost input states without trivially braiding them. The end result is shown in Figure \ref{Measurementbraid}, where we suppress the factor $\delta^{-2}$.
Note that in contrast to the topological moves that we have used, an algebraic approach based on expanding the braid into a sum of the basis elements leads to a complicated computation for Figure \ref{Measurementbraid}.

The entanglement-swapping protocol with braids holds for arbitrary $d$. In Figure \ref{Entanglement-Swapping} we illustrate the fermionic case $d=2$.
\begin{center}
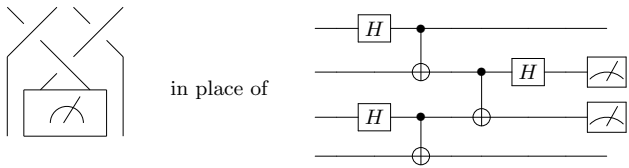
\begin{figure}[h]
\scalebox{0.88}{
 \begin{tikzpicture}
\draw (1.25,2.5) --(1,2.75);
\draw (2.25,1.5) --(1.5,2.25);
\draw (1,2) --(1.75,2.75);
\draw (2.25,2.5) --(2,2.75);
\draw (2.75,2) --(2.5,2.25);
\draw (2,2) --(2.75,2.75);
\draw (1.5,1.5) --(1.75,1.75);
\draw (1,0.8) --(1,2);
\draw (2.75,0.8) --(2.75,2);
\draw (2.5,1.5) --(1.25,1.5);
\draw (1.25,1.5) --(1.25,0.8);
\draw (2.5,1.5) --(2.5,0.8);
\draw (2.5,0.8) --(1.25,0.8);
\draw (2.15,1) arc (0:180:.25);
\draw (1.85,1) --(2.1,1.4);
\node  (3.75,0.35) at (4.2,1.5) {in place of};
\node  (7,0.35) at (8,1.5) {\Qcircuit @C=1em @R=.7em {
& \qw & \gate{H} & \ctrl{1} & \qw & \qw & \qw & \qw & \qw \\
& \qw & \qw & \targ&  \qw & \ctrl{1} & \gate{H} & \qw & \meter \\
& \qw & \gate{H} & \ctrl{1} & \qw & \targ & \qw & \qw & \meter \\
& \qw & \qw & \targ & \qw & \qw & \qw & \qw & \qw}};
\end{tikzpicture}} 
\caption{Entanglement-swapping with braids on the left, and a quantum circuit on the right.
\label{Entanglement-Swapping}}
\end{figure} 
\end{center}

\begin{remark}
Pictorial representation of other protocols, such as teleportation, superdense coding, and the EPR protocol for quantum key distribution, could be studied by these methods.  
\end{remark}

\iffalse
\begin{figure}[h]
\centering \begin{tikzpicture}
\draw (1.25,2.5) --(1,2.75);
\draw (2.25,1.5) --(1.5,2.25);
\draw (1,2) --(1.75,2.75);
\draw (2.25,2.5) --(2,2.75);
\draw (2.75,2) --(2.5,2.25);
\draw (2,2) --(2.75,2.75);
\draw (1.5,1.5) --(1.75,1.75);
\draw (1,0.8) --(1,2);
\draw (2.75,0.8) --(2.75,2);
\draw (2.5,1.5) --(1.25,1.5);
\draw (1.25,1.5) --(1.25,0.8);
\draw (2.5,1.5) --(2.5,0.8);
\draw (2.5,0.8) --(1.25,0.8);
\draw (2.15,1) arc (0:180:.25);
\draw (1.85,1) --(2.1,1.4);
\node  (3.75,0.35) at (4.75,1.5) {in place of};
\node  (7,0.35) at (9.5,1.5) {\Qcircuit @C=1em @R=.7em {
& \qw & \gate{H} & \ctrl{1} & \qw & \qw & \qw & \qw & \qw \\
& \qw & \qw & \targ&  \qw & \ctrl{1} & \gate{H} & \qw & \meter \\
& \qw & \gate{H} & \ctrl{1} & \qw & \targ & \qw & \qw & \meter \\
& \qw & \qw & \targ & \qw & \qw & \qw & \qw & \qw}};
\end{tikzpicture} \centering
\caption{Entanglement-swapping with braids on the left, and a quantum circuit on the right.
\label{Entanglement-Swapping}}
\end{figure} 
\fi

\section{Different models for quantum information}\label{sec:OtherModels}
In this section we introduce and contrast one-string, two-string, and four-string models, in order to represent quantum information in terms of diagrams.  Each of these different models has its own advantages in describing different applications. For example, see \S\ref{Protocols} for our use of the one-string model.  This model can also be imbedded into the two-string model.  In \S\ref{Sect:ControlledGates} we give an application of the four-string model.  

The qudits are given by parafermions in the one-string or two-string models. In the four-string model a single qudit is a particle-antiparticle pair, so it always has total charge zero. In both the two-string model and in the four-string model, the qudit Pauli $X,Y,Z$ matrices acting on 1-qudits can be represented by diagrams, see~\cite{JL}.  Here we give the algebraic form of the four-string representation in detail in  \S\ref{XYZ Matrices}.  

\subsection{The One-String Model}\label{Sect:DifferentModels-1}
In \S\ref{Notation} we realized a qudit by a single labeled string. We replace $m$-qudits, represented algebraically as  $|k_{1}k_{2} \cdots k_{m} \rangle$, by the diagram
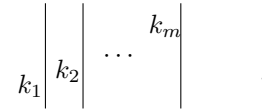
\[
 \begin{tikzpicture}
\node (0,0) at (2,-0.4) {$k_{1}$};
\draw (2.2,-0.7) --(2.2,0.7);
\node (0,0) at (2.5,-0.2) {$k_{2}$};
\draw (2.7,-0.7) --(2.7,0.7);
\node (0,0) at (3.2,0) {$\cdots$};
\node (0,0) at (3.8,0.4) {$k_{m}$};
\draw (4,-0.7) --(4,0.7);
\node (0,0) at (5.1,-0.3) {$.$};
\end{tikzpicture} 
\]
Transformations $T_{m}$ on $m$-qudits were realized by diagrams with $m$ input strings and $m$ output strings.  We represent them as a box with the $m$ input strings (on the top) and the $m$ output strings (on the bottom),  
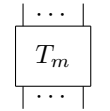
\[
\begin{tikzpicture}
\draw (-3.3,0.4) --(-2.3,0.4);
\draw (-3.3,-0.4) --(-3.3,0.4);
\draw (-3.3,-0.4) --(-2.3,-0.4);
\draw (-3.2,0.4) --(-3.2,0.7);
\draw (-3.2,-0.4) --(-3.2,-0.7);
\node (0,0) at (-2.8,0.55) {$\cdots$};
\node (0,0) at (-2.8,-0.55) {$\cdots$};
\node (0,0) at (-2.8,0) {$T_{m}$};
\draw (-2.4,0.4) --(-2.4,0.7);
\draw (-2.4,-0.4) --(-2.4,-0.7);
\draw (-2.3,-0.4) --(-2.3,0.4);
\node (0,0) at (-2,-0.3) {$.$};
\end{tikzpicture}
\]
The measurement on the space of $m$-qudits is represented by the trace, 
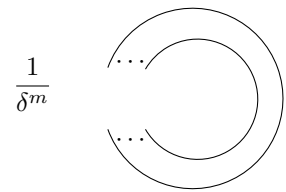
\[
\begin{tikzpicture}
\node (3.75,0.35) at (1,0) {$\mbox{$\displaystyle \frac{1}{\delta^{m}}$}$};
\draw (2.5,-0.6) arc (-150:150:0.8);
\draw (2,-0.6) arc (-160:160:1.2);
\node (0,0) at (2.15,0.3) {$\cdot$};
\node (0,0) at (2.30,0.3) {$\cdot$};
\node (0,0) at (2.45,0.3) {$\cdot$};
\node (0,0) at (2.15,-0.75) {$\cdot$};
\node (0,0) at (2.30,-0.75) {$\cdot$};
\node (0,0) at (2.45,-0.75) {$\cdot$};
\node  (3.75,0.35) at (5.3,-.5) {$.$};
\end{tikzpicture}
\]
We call this representation of quantum information the \textit{one-string model}. In this model the Hilbert space is $\mathbb{Z}_{d}$ graded. So, the transformations act on different components as the graded tensor product. We note that the diagrams used in the previous sections, such as the braid, are zero graded, or globally gauge invariant. Thus, the twisting scalar $q^{\alpha \beta}=1$, which reduces the graded tensor product to the usual tensor product.

\subsection{The Type I, Two-String Model}
In the \textit{type I, two-string model}, we realize a qudit by one labeled cap
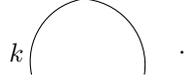
\begin{center}
\begin{tikzpicture}
\node  (3.75,0.35) at (-0.4,0.3) {$k$};
\coordinate (A) at (-0.2,0);
\coordinate (B) at (0.45,1);
\draw[color=black] (A) to [bend left=45] (B);
\draw (0.45,1) --(0.55,1);
\coordinate (A) at (0.55,1);
\coordinate (B) at (1.3,0);
\draw[color=black] (A) to [bend left=45] (B);
\node  (3.75,0.35) at (1.8,0.3) {$.$};
\end{tikzpicture}
\end{center}
Here $0 \leq k \leq d-1$. We represent $m$-qudits by
\smallskip
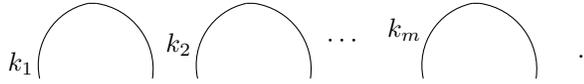
\begin{center}
\begin{tikzpicture}
\node  (3.75,0.35) at (-0.45,0.2) {$k_{1}$};
\coordinate (A) at (-0.2,0);
\coordinate (B) at (0.45,1);
\draw[color=black] (A) to [bend left=45] (B);
\draw (0.45,1) --(0.55,1);
\coordinate (A) at (0.55,1);
\coordinate (B) at (1.3,0);
\draw[color=black] (A) to [bend left=45] (B);
\node  (3.75,0.35) at (1.65,0.45) {$k_{2}$};
\coordinate (C) at (1.9,0);
\coordinate (D) at (2.55,1);
\draw[color=black] (C) to [bend left=45] (D);
\draw (2.55,1) --(2.65,1);
\coordinate (C) at (2.65,1);
\coordinate (D) at (3.4,0);
\draw[color=black] (C) to [bend left=45] (D);
\node  (3.75,0.35) at (3.85,0.5) {$\cdots$};
\node  (3.75,0.35) at (4.65,0.65) {$k_{m}$};
\coordinate (E) at (4.9,0);
\coordinate (F) at (5.55,1);
\draw[color=black] (E) to [bend left=45] (F);
\draw (5.55,1) --(5.65,1);
\coordinate (E) at (5.65,1);
\coordinate (F) at (6.4,0);
\draw[color=black] (E) to [bend left=45] (F);
\node  (3.75,0.35) at (7,0.3) {$.$};
\end{tikzpicture}
\end{center}

We represent a transformation $T_{2m}$ on $m$-qudits by a box with $2m$ input strings on top and $2m$ output strings on the bottom,
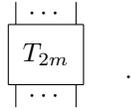
\begin{center}
\begin{tikzpicture}
\draw (-3.3,0.4) --(-2.3,0.4);
\draw (-3.3,-0.4) --(-3.3,0.4);
\draw (-3.3,-0.4) --(-2.3,-0.4);
\draw (-3.2,0.4) --(-3.2,0.7);
\draw (-3.2,-0.4) --(-3.2,-0.7);
\node (0,0) at (-2.8,0.55) {$\cdots$};
\node (0,0) at (-2.8,-0.55) {$\cdots$};
\node (0,0) at (-2.8,0) {$T_{2m}$};
\draw (-2.4,0.4) --(-2.4,0.7);
\draw (-2.4,-0.4) --(-2.4,-0.7);
\draw (-2.3,-0.4) --(-2.3,0.4);
\node  (3.75,0.35) at (-1.7,-0.3) {$.$};
\end{tikzpicture}
\end{center}

The one-string model can be embedded into the type I two-string model by making the following replacements:
\begin{equation*}
\scalebox{0.9}{
\begin{tikzpicture}
\node (3.75,0.35) at (-4,0.5) {$\mbox{Qudit}:$};
\node  (3.75,0.35) at (-2.2,0.5) {$k$};
\draw (-2,0) --(-2,1);
\node  (3.75,0.35) at (-1.2,0.5) {$\rightarrow$};
\node  (3.75,0.35) at (-0.4,0.3) {$k$};
\coordinate (A) at (-0.2,0);
\coordinate (B) at (0.45,1);
\draw[color=black] (A) to [bend left=45] (B);
\draw (0.45,1) --(0.55,1);
\coordinate (A) at (0.55,1);
\coordinate (B) at (1.3,0);
\draw[color=black] (A) to [bend left=45] (B);
\node  (3.75,0.35) at (1.7,0.2) {$,$};
\node  (0,0) at (7.5,0.2) {$$};
\end{tikzpicture}}
\end{equation*}
\begin{equation*}
\scalebox{0.75}{
\begin{tikzpicture}
\node (3.75,0.35) at (1,0) {$\mbox{\large $m$ Qudits}:$};
\node (0,0) at (2.5,-0.2) {$k_{1}$};
\draw (2.7,-0.7) --(2.7,0.7);
\node (0,0) at (3.3,0) {$\cdots$};
\node (0,0) at (3.7,0.4) {$k_{m}$};
\draw (4,-0.7) --(4,0.7);
\node (3.75,0.35) at (4.8,0) {$\rightarrow$};
\node (3.75,0.35) at (5.6,-0.2) {$k_{1}$};
\coordinate (A) at (5.85,-0.4);
\coordinate (B) at (6.5,0.6);
\draw[color=black] (A) to [bend left=45] (B);
\draw (6.5,0.6) --(6.6,0.6);
\coordinate (A) at (6.6,0.6);
\coordinate (B) at (7.25,-0.4);
\draw[color=black] (A) to [bend left=45] (B);
\node (3.75,0.35) at (7.9,0) {$\cdots$};
\node  (3.75,0.35) at (8.5,0.2) {$k_{m}$};
\coordinate (A) at (8.75,-0.4);
\coordinate (B) at (9.4,0.6);
\draw[color=black] (A) to [bend left=45] (B);
\draw (9.4,0.6) --(9.5,0.6);
\coordinate (A) at (9.5,0.6);
\coordinate (B) at (10.15,-0.4);
\draw[color=black] (A) to [bend left=45] (B);
\node  (3.75,0.35) at (10.5,-0.2) {$,$};
\node  (3.75,0.35) at (11.5,-0.2) {$$};
\end{tikzpicture}}
\end{equation*} 
\begin{equation*}
\scalebox{0.85}{
\begin{tikzpicture}
\node (3.75,0.35) at (-8.2,0) {$\mbox{Transformation}:$};
\draw (-6.3,0.4) --(-5.3,0.4);
\draw (-6.3,-0.4) --(-6.3,0.4);
\draw (-6.3,-0.4) --(-5.3,-0.4);
\draw (-6.2,0.4) --(-6.2,0.7);
\draw (-6.2,-0.4) --(-6.2,-0.7);
\node (0,0) at (-5.8,0.55) {$\cdots$};
\node (0,0) at (-5.8,-0.55) {$\cdots$};
\node (0,0) at (-5.8,0) {$T_{m}$};
\draw (-5.4,0.4) --(-5.4,0.7);
\draw (-5.4,-0.4) --(-5.4,-0.7);
\draw (-5.3,-0.4) --(-5.3,0.4);
\node  (3.75,0.35) at (-4.6,0) {$\rightarrow$};
\draw (-3.6,-0.4) --(-3.6,0.4);
\draw (-3.6,0.4) --(-3.4,0.4);
\draw (-3.6,-0.4) --(-3.4,-0.4);
\draw (-3.6,-0.4) --(-3.4,-0.4);
\draw (-3.5,0.4) --(-3.5,0.7);
\draw (-3.5,-0.4) --(-3.5,-0.7);
\draw (-3.2,-0.4) --(-2.6,-0.4);
\draw (-3.2,0.4) --(-2.6,0.4);
\draw (-3.3,-0.7) --(-3.3,0.7);
\node (0.1,0.5) at (-3,0.5) {$\cdots$};
\node (0.1,-0.6) at (-3,-0.6) {$\cdots$};
\draw (-2.7,0.4) --(-2.7,0.7);
\draw (-2.7,-0.4) --(-2.7,-0.7);
\draw (-2.5,-0.7) --(-2.5,0.7);
\draw (-2.4,0.4) --(-2.3,0.4);
\draw (-2.4,-0.4) --(-2.3,-0.4);
\draw (-2.3,-0.4) --(-2.3,0.4);
\node (0,0) at (-3,0) {$T_{m}$};
\node (0.1,-0.6) at (-2,-0.3) {$,$};
\node (0.1,-0.6) at (3.4,-0.6) {$$};
\end{tikzpicture}}
\label{Transformation}
\end{equation*}
\begin{equation*}
\scalebox{0.72}{
\begin{tikzpicture}
\node (3.75,0.35) at (0.5,0) {$\mbox{\large Measurement}:$};
\draw (2.5,-0.6) arc (-150:150:0.8);
\draw (2,-0.6) arc (-160:160:1.2);
\node (0,0) at (2.15,0.3) {$\cdot$};
\node (0,0) at (2.30,0.3) {$\cdot$};
\node (0,0) at (2.45,0.3) {$\cdot$};
\node (0,0) at (2.15,-0.75) {$\cdot$};
\node (0,0) at (2.30,-0.75) {$\cdot$};
\node (0,0) at (2.45,-0.75) {$\cdot$};
\node  (3.75,0.35) at (4.8,-0.3) {$\rightarrow$};
\coordinate (A) at (5.5,0.2);
\coordinate (B) at (6.25,-0.8);
\draw[color=black] (A) to [bend left=-45] (B);
\coordinate (A) at (6.25,-0.8);
\coordinate (B) at (6.35,-0.8);
\draw[color=black] (A) to [bend left=-5] (B);
\coordinate (A) at (6.35,-0.8);
\coordinate (B) at (7.1,0.2);
\draw[color=black] (A) to [bend left=-45] (B);
\node  (3.75,0.35) at (7.7,-0.3) {$\cdots$};
\coordinate (A) at (8.3,0.2);
\coordinate (B) at (9.05,-0.8);
\draw[color=black] (A) to [bend left=-45] (B);
\coordinate (A) at (9.05,-0.8);
\coordinate (B) at (9.15,-0.8);
\draw[color=black] (A) to [bend left=-5] (B);
\coordinate (A) at (9.15,-0.8);
\coordinate (B) at (9.9,0.2);
\draw[color=black] (A) to [bend left=-45] (B);
\node  (3.75,0.35) at (10.1,-0.5) {$.$};
\node (0,0) at (10.6,0) {$$};
\end{tikzpicture}}
\end{equation*} 

\subsection{The Type II, Two-String Model}
In the type II, two-string model, we represent $m$-qudits by labeled caps
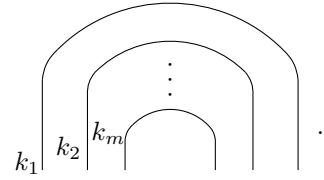
\begin{center}
\begin{tikzpicture}
\coordinate (A) at (-0.5,0.6);
\coordinate (B) at (0.5,0.6);
\draw[color=black] (A) to [bend left=45] (B);
\coordinate (A) at (-0.5,0.6);
\coordinate (B) at (-0.6,0.4);
\draw[color=black] (A) to [bend left=-20] (B);
\draw (-0.6,0) --(-0.6,0.4);
\coordinate (A) at (0.5,0.6);
\coordinate (B) at (0.6,0.4);
\draw[color=black] (A) to [bend left=20] (B);
\draw (0.6,0) --(0.6,0.4);
\node  (3.75,0.35) at (-0.82,0.5) {$k_{m}$};
\coordinate (A) at (-1,1.3);
\coordinate (B) at (1,1.3);
\draw[color=black] (A) to [bend left=45] (B);
\coordinate (A) at (-1,1.3);
\coordinate (B) at (-1.1,1);
\draw[color=black] (A) to [bend left=-20] (B);
\coordinate (A) at (1,1.3);
\coordinate (B) at (1.1,1);
\draw[color=black] (A) to [bend left=20] (B);
\draw (-1.1,0) --(-1.1,1);
\draw (1.1,0) --(1.1,1);
\node  (3.75,0.35) at (-1.35,0.3) {$k_{2}$};
\node  (3.75,0.35) at (0,1.4) {$\cdot$};
\node  (3.75,0.35) at (0,1.2) {$\cdot$};
\node  (3.75,0.35) at (0,1) {$\cdot$};
\coordinate (A) at (-1.5,1.6);
\coordinate (B) at (1.5,1.6);
\draw[color=black] (A) to [bend left=45] (B);
\coordinate (A) at (-1.5,1.6);
\coordinate (B) at (-1.7,1.2);
\draw[color=black] (A) to [bend left=-20] (B);
\coordinate (A) at (1.5,1.6);
\coordinate (B) at (1.7,1.2);
\draw[color=black] (A) to [bend left=20] (B);
\draw (1.7,0) --(1.7,1.2);
\draw (-1.7,0) --(-1.7,1.2);
\node  (3.75,0.35) at (-1.9,0.1) {$k_{1}$};
\node  (3.75,0.35) at (2,0.5) {$.$};
\end{tikzpicture}
\end{center}

In this model we represent a transformation $T_{2m}$ on $m$-qudits by a box with $2m$ input strings on top and $2m$ output strings on the bottom (as in the type I model), namely
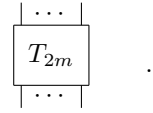
\begin{center}
\begin{tikzpicture}
\draw (-3.3,0.4) --(-2.3,0.4);
\draw (-3.3,-0.4) --(-3.3,0.4);
\draw (-3.3,-0.4) --(-2.3,-0.4);
\draw (-3.2,0.4) --(-3.2,0.7);
\draw (-3.2,-0.4) --(-3.2,-0.7);
\node (0,0) at (-2.8,0.55) {$\cdots$};
\node (0,0) at (-2.8,-0.55) {$\cdots$};
\node (0,0) at (-2.8,0) {$T_{2m}$};
\draw (-2.4,0.4) --(-2.4,0.7);
\draw (-2.4,-0.4) --(-2.4,-0.7);
\draw (-2.3,-0.4) --(-2.3,0.4);
\node  (3.75,0.35) at (-2,0.6) {$$};
\node  (3.75,0.35) at (-2,-0.6) {$$};
\node  (3.75,0.35) at (-1.5,-0.2) {$.$};
\end{tikzpicture}
\end{center}

The one-string model can be embedded into the type II two-string model by replacements of a transformation similar to type I model. The image of the transformation acting on $m$-qudits becomes
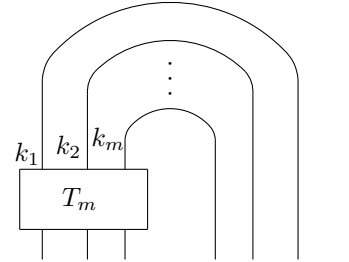
\begin{center}
\begin{equation*}
\begin{tikzpicture}
\coordinate (A) at (-0.5,0.6);
\coordinate (B) at (0.5,0.6);
\draw[color=black] (A) to [bend left=45] (B);
\coordinate (A) at (-0.5,0.6);
\coordinate (B) at (-0.6,0.4);
\draw[color=black] (A) to [bend left=-20] (B);
\draw (-0.6,0) --(-0.6,0.4);
\coordinate (A) at (0.5,0.6);
\coordinate (B) at (0.6,0.4);
\draw[color=black] (A) to [bend left=20] (B);
\draw (0.6,-1.2) --(0.6,0.4);
\node  (3.75,0.35) at (-0.82,0.4) {$k_{m}$};
\coordinate (A) at (-1,1.3);
\coordinate (B) at (1,1.3);
\draw[color=black] (A) to [bend left=45] (B);
\coordinate (A) at (-1,1.3);
\coordinate (B) at (-1.1,1);
\draw[color=black] (A) to [bend left=-20] (B);
\coordinate (A) at (1,1.3);
\coordinate (B) at (1.1,1);
\draw[color=black] (A) to [bend left=20] (B);
\draw (-1.1,0) --(-1.1,1);
\draw (1.1,-1.2) --(1.1,1);
\node  (3.75,0.35) at (-1.35,0.3) {$k_{2}$};
\node  (3.75,0.35) at (0,1.4) {$\cdot$};
\node  (3.75,0.35) at (0,1.2) {$\cdot$};
\node  (3.75,0.35) at (0,1) {$\cdot$};
\coordinate (A) at (-1.5,1.6);
\coordinate (B) at (1.5,1.6);
\draw[color=black] (A) to [bend left=45] (B);
\coordinate (A) at (-1.5,1.6);
\coordinate (B) at (-1.7,1.2);
\draw[color=black] (A) to [bend left=-20] (B);
\coordinate (A) at (1.5,1.6);
\coordinate (B) at (1.7,1.2);
\draw[color=black] (A) to [bend left=20] (B);
\draw (1.7,-1.2) --(1.7,1.2);
\draw (-1.7,0) --(-1.7,1.2);
\node  (3.75,0.35) at (-1.9,0.2) {$k_{1}$};
\draw (-0.3,0) --(-2,0);
\draw (-0.3,-0.8) --(-2,-0.8);
\draw (-0.3,-0.8) --(-0.3,0);
\draw (-2,-0.8) --(-2,0);
\node  (3.75,0.35) at (-1.2,-.4) {$T_{m}$};
\draw (-0.6,-0.8) --(-0.6,-1.2);
\draw (-1.1,-0.8) --(-1.1,-1.2);
\draw (-1.7,-0.8) --(-1.7,-1.2);
\node  (3.75,0.35) at (2.4,-0.2) {$.$};
\end{tikzpicture}
\end{equation*}
\end{center}

\noindent The measurement is represented by
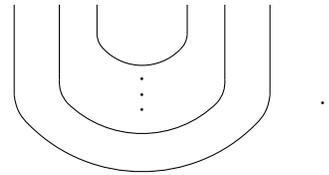
\begin{center}
\begin{equation}
\begin{tikzpicture}
\coordinate (A) at (-0.5,-0.6);
\coordinate (B) at (0.5,-0.6);
\draw[color=black] (A) to [bend left=-45] (B);
\coordinate (A) at (-0.5,-0.6);
\coordinate (B) at (-0.6,-0.4);
\draw[color=black] (A) to [bend left=20] (B);
\draw (-0.6,0) --(-0.6,-0.4);
\coordinate (A) at (0.5,-0.6);
\coordinate (B) at (0.6,-0.4);
\draw[color=black] (A) to [bend left=-20] (B);
\draw (0.6,0) --(0.6,-0.4);
%\node  (3.75,0.35) at (0.4,-0.2) {$k_{1}$};
\coordinate (A) at (-1,-1.3);
\coordinate (B) at (1,-1.3);
\draw[color=black] (A) to [bend left=-45] (B);
\coordinate (A) at (-1,-1.3);
\coordinate (B) at (-1.1,-1);
\draw[color=black] (A) to [bend left=20] (B);
\coordinate (A) at (1,-1.3);
\coordinate (B) at (1.1,-1);
\draw[color=black] (A) to [bend left=-20] (B);
\draw (-1.1,0) --(-1.1,-1);
\draw (1.1,0) --(1.1,-1);
\node  (3.75,0.35) at (0,-1.4) {$\cdot$};
\node  (3.75,0.35) at (0,-1.2) {$\cdot$};
\node  (3.75,0.35) at (0,-1) {$\cdot$};
\coordinate (A) at (-1.5,-1.6);
\coordinate (B) at (1.5,-1.6);
\draw[color=black] (A) to [bend left=-45] (B);
\coordinate (A) at (-1.5,-1.6);
\coordinate (B) at (-1.7,-1.2);
\draw[color=black] (A) to [bend left=20] (B);
\coordinate (A) at (1.5,-1.6);
\coordinate (B) at (1.7,-1.2);
\draw[color=black] (A) to [bend left=-20] (B);
\draw (1.7,0) --(1.7,-1.2);
\draw (-1.7,0) --(-1.7,-1.2);
\node  (3.75,0.35) at (2.4,-1.3) {$.$};
\end{tikzpicture}
\end{equation}
\end{center}

\subsection{The Four-String Model\label{Sect:DifferentModels-4}}
For the type I four-string model, we realize a qudit  by two labeled caps. We represent $m$-qudits by the picture:
\begin{center}
\scalebox{0.68}{
\begin{tikzpicture}
\node  (3.75,0.35) at (-0.7,0.3) {$k_{1}$};
\coordinate (A) at (-0.4,0);
\coordinate (B) at (0.25,1);
\draw[color=black] (A) to [bend left=50] (B);
\draw (0.25,1) --(0.35,1);
\coordinate (A) at (0.35,1);
\coordinate (B) at (1,0);
\draw[color=black] (A) to [bend left=50] (B);
\node  (3.75,0.35) at (2.2,0.3) {$-k_{1}$};
\node  (3.75,0.35) at (3.2,0.3) {$k_{2}$};
\coordinate (A) at (1.2,0);
\coordinate (B) at (1.85,1);
\draw[color=black] (A) to [bend left=50] (B);
\draw (1.85,1) --(1.95,1);
\coordinate (A) at (1.95,1);
\coordinate (B) at (2.6,0);
\draw[color=black] (A) to [bend left=50] (B);
\coordinate (A) at (3.5,0);
\coordinate (B) at (4.15,1);
\draw[color=black] (A) to [bend left=50] (B);
\draw (4.15,1) --(4.25,1);
\coordinate (A) at (4.25,1);
\coordinate (B) at (4.9,0);
\draw[color=black] (A) to [bend left=50] (B);
\node  (3.75,0.35) at (6.1,0.3) {$-k_{2}$};
\coordinate (A) at (5.1,0);
\coordinate (B) at (5.75,1);
\draw[color=black] (A) to [bend left=45] (B);
\draw (5.75,1) --(5.85,1);
\coordinate (A) at (5.85,1);
\coordinate (B) at (6.5,0);
\draw[color=black] (A) to [bend left=45] (B);
\node  (3.75,0.35) at (7,0.4) {$\cdots$};
\node  (3.75,0.35) at (7.7,0.3) {$k_{m}$};
\coordinate (A) at (8,0);
\coordinate (B) at (8.65,1);
\draw[color=black] (A) to [bend left=50] (B);
\draw (8.65,1) --(8.75,1);
\coordinate (A) at (8.75,1);
\coordinate (B) at (9.4,0);
\draw[color=black] (A) to [bend left=50] (B);
\node  (3.75,0.35) at (10.6,0.3) {$-k_{m}$};
\coordinate (A) at (9.6,0);
\coordinate (B) at (10.25,1);
\draw[color=black] (A) to [bend left=50] (B);
\draw (10.25,1) --(10.35,1);
\coordinate (A) at (10.35,1);
\coordinate (B) at (11,0);
\draw[color=black] (A) to [bend left=50] (B);
\node  (3.75,0.35) at (11.4,0) {$.$};
\end{tikzpicture}}
\end{center}
\noindent Here $0 \leq k_{i} \leq d-1$. 
For the type II four-string model, we represent $m$-qudits by the diagram:
\begin{center}
\scalebox{0.75}{
\begin{tikzpicture}
\coordinate (A) at (-0.5,0.6);
\coordinate (B) at (0.5,0.6);
\draw[color=black] (A) to [bend left=45] (B);
\coordinate (A) at (-0.5,0.6);
\coordinate (B) at (-0.6,0.4);
\draw[color=black] (A) to [bend left=-20] (B);
\draw (-0.6,0) --(-0.6,0.4);
\coordinate (A) at (0.5,0.6);
\coordinate (B) at (0.6,0.4);
\draw[color=black] (A) to [bend left=20] (B);
\draw (0.6,0) --(0.6,0.4);
\coordinate (A) at (-1,0.9);
\coordinate (B) at (1,0.9);
\draw[color=black] (A) to [bend left=30] (B);
\coordinate (A) at (-1,0.9);
\coordinate (B) at (-1.35,0.5);
\draw[color=black] (A) to [bend left=-20] (B);
\coordinate (A) at (1,0.9);
\coordinate (B) at (1.35,0.5);
\draw[color=black] (A) to [bend left=20] (B);
\draw (-1.35,0) --(-1.35,0.5);
\draw (1.35,0) --(1.35,0.5);
\node  (3.75,0.35) at (-0.95,0.3) {$-k_{1}$};
\node  (3.75,0.35) at (-1.55,0.3) {$k_{1}$};
\node  (3.75,0.35) at (2.4,0.3) {$-k_{2}$};
\node  (3.75,0.35) at (1.8,0.3) {$k_{2}$};
\coordinate (A) at (2.35,0.9);
\coordinate (B) at (2,0.5);
\draw[color=black] (A) to [bend left=-20] (B);
\coordinate (A) at (2.35,0.9);
\coordinate (B) at (4.35,0.9);
\draw[color=black] (A) to [bend left=30] (B);
\coordinate (A) at (4.35,0.9);
\coordinate (B) at (4.7,0.5);
\draw[color=black] (A) to [bend left=20] (B);
\draw (2,0) --(2,0.5);
\draw (4.7,0) --(4.7,0.5);
\coordinate (A) at (2.85,0.6);
\coordinate (B) at (3.85,0.6);
\draw[color=black] (A) to [bend left=45] (B);
\coordinate (A) at (3.85,0.6);
\coordinate (B) at (3.95,0.4);
\draw[color=black] (A) to [bend left=20] (B);
\coordinate (A) at (2.85,0.6);
\coordinate (B) at (2.75,0.4);
\draw[color=black] (A) to [bend left=-20] (B);
\draw (2.75,0) --(2.75,0.4);
\draw (3.95,0) --(3.95,0.4);
\node  (3.75,0.35) at (5.3,0.4) {$\cdots$};
\node  (3.75,0.35) at (6.72,0.3) {$- k_{m}$};
\node  (3.75,0.35) at (6,0.3) {$k_{m}$};
\coordinate (A) at (6.65,0.9);
\coordinate (B) at (6.3,0.5);
\draw[color=black] (A) to [bend left=-20] (B);
\coordinate (A) at (6.65,0.9);
\coordinate (B) at (6.65,0.9);
\draw[color=black] (A) to [bend left=30] (B);
\draw[color=black] (A) to [bend left=-20] (B);
\coordinate (A) at (6.65,0.9);
\coordinate (B) at (8.65,0.9);
\draw[color=black] (A) to [bend left=30] (B);
\coordinate (A) at (8.65,0.9);
\coordinate (B) at (8.95,0.5);
\draw[color=black] (A) to [bend left=20] (B);
\draw (6.3,0) --(6.3,0.5);
\draw (8.95,0) --(8.95,0.5);
\coordinate (A) at (7.3,0.6);
\coordinate (B) at (8.1,0.6);
\draw[color=black] (A) to [bend left=45] (B);
\coordinate (A) at (7.3,0.6);
\coordinate (B) at (7.2,0.4);
\draw[color=black] (A) to [bend left=-20] (B);
\coordinate (A) at (8.1,0.6);
\coordinate (B) at (8.2,0.4);
\draw[color=black] (A) to [bend left=20] (B);
\draw (7.2,0) --(7.2,0.4);
\draw (8.2,0) --(8.2,0.4);
\node  (3.75,0.35) at (9.3,0) {$.$};
\end{tikzpicture}}
\end{center}
\noindent Again $0 \leq k_{i} \leq d-1$. 

We represent a transformation $T_{4m}$ on $m$-qudits by a box with $4m$ input strings on top and $4m$ output strings on bottom,
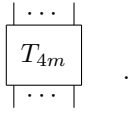
\begin{center}
\begin{tikzpicture}
\draw (-3.3,0.4) --(-2.3,0.4);
\draw (-3.3,-0.4) --(-3.3,0.4);
\draw (-3.3,-0.4) --(-2.3,-0.4);
\draw (-3.2,0.4) --(-3.2,0.7);
\draw (-3.2,-0.4) --(-3.2,-0.7);
\node (0,0) at (-2.8,0.55) {$\cdots$};
\node (0,0) at (-2.8,-0.55) {$\cdots$};
\node (0,0) at (-2.8,0) {$T_{4m}$};
\draw (-2.4,0.4) --(-2.4,0.7);
\draw (-2.4,-0.4) --(-2.4,-0.7);
\draw (-2.3,-0.4) --(-2.3,0.4);
\node  (3.75,0.35) at (-1.7,-0.3) {$.$};
\end{tikzpicture}
\end{center}

%\begin{center}
%\begin{tikzpicture}
%\draw (-3.3,0.4) --(-2.3,0.4);
%\draw (-3.3,-0.4) --(-3.3,0.4);
%\draw (-3.3,-0.4) --(-2.3,-0.4);
%\draw (-3.2,0.4) --(-3.2,0.7);
%\draw (-2.95,0.4) --(-2.95,0.7);
%\draw (-2.95,-0.4) --(-2.95,-0.7);
%\draw (-2.7,0.4) --(-2.7,0.7);
%\draw (-2.7,-0.4) --(-2.7,-0.7);
%\draw (-3.2,-0.4) --(-3.2,-0.7);
%\node (0,0) at (-2.8,0) {$X$};
%\draw (-2.4,0.4) --(-2.4,0.7);
%\draw (-2.4,-0.4) --(-2.4,-0.7);
%\draw (-2.3,-0.4) --(-2.3,0.4);
%\node  (3.75,0.35) at (-2,-0.2) {$.$};
%\end{tikzpicture}
%\end{center}

Note that the four-string model is different from the one-string model and the two-string model, where all operators in the parafermion algebra with the proper number of generators  preserve the space of qudits, namely diagrams given by caps. 
For instance, the following diagram does not preserve the space of $2$-qudits:
\medskip
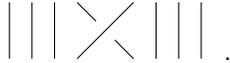
\begin{center} 
\begin{tikzpicture}
\draw (0.1,2) --(0.1,2.75);
\draw (0.4,2) --(0.4,2.75);
\draw (0.7,2) --(0.7,2.75);
\draw (1.25,2.5) --(1,2.75);
\draw (1.5,2.25) --(1.75,2);
\draw (1,2) --(1.75,2.75);
\draw (2.05,2) --(2.05,2.75);
\draw (2.35,2) --(2.35,2.75);
\draw (2.65,2) --(2.65,2.75);
\node  (3.75,0.35) at (3,2) {$.$};
\end{tikzpicture}
\end{center}

In the four-string model we are interested in diagrams that do preserve the space spanned by qudits. An operator in the parafermion algebra with $4m$ generators preserves the space of $m$-qudits if and only if it is 
of the following form, 
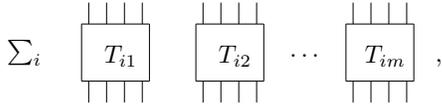
\begin{center}
\scalebox{0.95}{
\begin{tikzpicture}
\node  (3.75,0.35) at (-4.1,0) {$\sum_{i}$};
\draw (-3.3,0.4) --(-2.35,0.4);
\draw (-3.3,-0.4) --(-3.3,0.4);
\draw (-3.3,-0.4) --(-2.35,-0.4);
\draw (-3.2,0.4) --(-3.2,0.7);
\draw (-3.2,-0.4) --(-3.2,-0.7);
\draw (-2.7,0.4) --(-2.7,0.7);
\draw (-2.7,-0.4) --(-2.7,-0.7);
\draw (-2.95,0.4) --(-2.95,0.7);
\draw (-2.95,-0.4) --(-2.95,-0.7);
\node (0,0) at (-2.75,-0.05) {$T_{i 1}$};
\draw (-2.45,0.4) --(-2.45,0.7);
\draw (-2.45,-0.4) --(-2.45,-0.7);
\draw (-2.35,-0.4) --(-2.35,0.4);
\draw (-1.7,0.4) --(-0.75,0.4);
\draw (-1.7,-0.4) --(-0.75,-0.4);
\draw (-1.7,-0.4) --(-1.7,0.4);
\draw (-0.75,-0.4) --(-0.75,0.4);
\node (0,0) at (-1.15,-0.05) {$T_{i 2}$};
\draw (-0.85,0.4) --(-0.85,0.7);
\draw (-0.85,-0.4) --(-0.85,-0.7);
\draw (-1.1,0.4) --(-1.1,0.7);
\draw (-1.1,-0.4) --(-1.1,-0.7);
\draw (-1.35,0.4) --(-1.35,0.7);
\draw (-1.35,-0.4) --(-1.35,-0.7);
\draw (-1.6,0.4) --(-1.6,0.7);
\draw (-1.6,-0.4) --(-1.6,-0.7);
\node (0,0) at (-0.15,-0.05) {$\cdots$};
\draw (0.4,0.4) --(1.35,0.4);
\draw (0.4,-0.4) --(0.4,0.4);
\draw (1.35,-0.4) --(1.35,0.4);
\draw (0.4,-0.4) --(1.35,-0.4);
\node (0,0) at (0.95,-0.05) {$T_{i m}$};
\draw (0.5,-0.4) --(0.5,-0.7);
\draw (0.5,0.4) --(0.5,0.7);
\draw (0.75,-0.4) --(0.75,-0.7);
\draw (0.75,0.4) --(0.75,0.7);
\draw (1,-0.4) --(1,-0.7);
\draw (1,0.4) --(1,0.7);
\draw (1.25,-0.4) --(1.25,-0.7);
\draw (1.25,0.4) --(1.25,0.7);
\node  (3.75,0.35) at (1.7,-0.15) {,};
%\node  (3.75,0.35) at (4,-0.05) {$\mbox{Here $X_{i j}$ is zero graded.}$};
\end{tikzpicture}}
\end{center}
where each $T_{i j}$ is zero graded.
A diagrammatic example is the double braid in Figure \ref{Double Braid} in \S \ref{Sect:ControlledGates}

\subsection{Double Braids as Controlled Gates}\label{Sect:ControlledGates}
Let us construct some controlled transformations for the type I four-string model.  
For a transformation $A$ acting on a single qudit, define the controlled transformation $C_A$ on 2-qudit states $\ket{ij}=\ket{i}\ket{j}$ as 
	\be
		C_{A} \ket{ij}
		= \ket{i} \,A^{i} \ket{j}\;.
	\ee
We use the matrices $X$, $Y$, $Z$  given in \eqref{xyz1} for $A$, to describe the action on a single qudit, giving $C_{X}, C_{Y}, C_{Z}$.

\begin{figure}[h]
\scalebox{1.05}{
\begin{tikzpicture}
\draw (-1.1,-4) --(-1.1,0.5);
\draw (2.9,-4) --(2.9,0.5);
\draw (-0.8,-4) --(-0.8,0.5);
\draw (-0.5,0) --(-0.5,0.5);
\draw (0.5,0) --(0.5,0.5);
\coordinate (A) at (0.75,-0.35);
\coordinate (B) at (0.5,0);
\draw[color=black] (A) to [bend left=45] (B);
\coordinate (A) at (-0.5,0);
\coordinate (B) at (0,-0.85);
\draw[color=black] (A) to [bend left=-30] (B);
\draw (1,0) --(1,0.5);
\draw (2.1,0) --(2.1,0.5);
\coordinate (A) at (1,0);
\coordinate (B) at (0,-1);
\draw[color=black] (A) to [bend left=25] (B);
\coordinate (A) at (-0.6,-1.8);
\coordinate (B) at (0,-1);
\draw[color=black] (B) to [bend left=-40] (A);
\coordinate (A) at (-0.6,-1.8);
\coordinate (B) at (0,-2.6);
\draw[color=black] (A) to [bend left=-40] (B);
\coordinate (A) at (1.5,-1);
\coordinate (B) at (2.1,0);
\draw[color=black] (B) to [bend left=25] (A);
\coordinate (A) at (1.5,-1);
\coordinate (B) at (0.75,-1.5);
\draw[color=black] (B) to [bend left=-5] (A);
\coordinate (A) at (0.4,-1.8);
\coordinate (B) at (0.75,-1.5);
\draw[color=black] (B) to [bend left=-30] (A);
\coordinate (A) at (0.4,-1.8);
\coordinate (B) at (0.75,-2.2);
\draw[color=black] (A) to [bend left=-40] (B);
\coordinate (A) at (1.15,-1.4);
\coordinate (B) at (1.5,-1.8);
\draw[color=black] (A) to [bend left=40] (B);
\coordinate (A) at (1.5,-1.8);
\coordinate (B) at (1.2,-2.5);
\draw[color=black] (A) to [bend left=25] (B);
\coordinate (A) at (1.2,-2.5);
\coordinate (B) at (0.8,-2.8);
\draw[color=black] (A) to [bend left=10] (B);
\coordinate (A) at (0,-3.3);
\coordinate (B) at (0.8,-2.8);
\draw[color=black] (B) to [bend left=3] (A);
\coordinate (A) at (1.85,-0.8);
\coordinate (B) at (2.25,-1.2);
\draw[color=black] (A) to [bend left=35] (B);
\coordinate (A) at (1.85,-2.85);
\coordinate (B) at (2.25,-1.2);
\draw[color=black] (B) to [bend left=35] (A);
\coordinate (A) at (1.85,-2.85);
\coordinate (B) at (1.5,-3.2);
\draw[color=black] (A) to [bend left=10] (B);
\coordinate (A) at (-0.5,-4);
\coordinate (B) at (0,-3.3);
\draw[color=black] (A) to [bend left=20] (B);
\coordinate (A) at (0.2,-4);
\coordinate (B) at (1.5,-3.2);
\draw[color=black] (B) to [bend left=2] (A);
\coordinate (A) at (0.5,-2.8);
\coordinate (B) at (0,-2.6);
\draw[color=black] (B) to [bend left=-10] (A);
\coordinate (A) at (1.12,-2.3);
\coordinate (B) at (0.75,-2.2);
\draw[color=black] (B) to [bend left=-5] (A);
\coordinate (A) at (1.7,-3.15);
\coordinate (B) at (1.6,-4);
\draw[color=black] (A) to [bend left=10] (B);
\coordinate (A) at (2.2,-2.6);
\coordinate (B) at (2.5,-2.95);
\draw[color=black] (A) to [bend left=10] (B);
\coordinate (A) at (2.5,-2.95);
\coordinate (B) at (2.55,-4);
\draw[color=black] (A) to [bend left=10] (B);
\draw (3.3,-4) --(3.3,0.5);
\end{tikzpicture}}
\caption{The Double Braid.  \label{Double Braid}}
\end{figure}
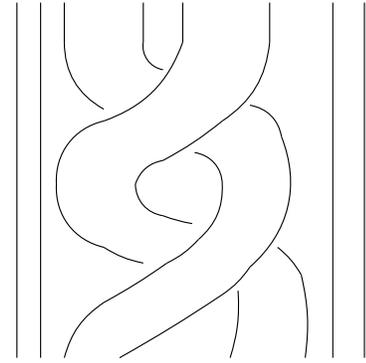

\begin{figure}[h]
\begin{center}
\begin{tikzpicture}
\node (3.75,0.35) at (-1.7,-1) {$$};
\node  (3.75,0.35) at (-0.85,0) {$-i$};
\node  (3.75,0.35) at (0.85,0.25) {$j$};
\coordinate (A) at (0,0.5);
\coordinate (B) at (-0.5,0);
\draw[color=black] (B) to [bend left=45] (A);
\coordinate (A) at (0,0.5);
\coordinate (B) at (0.5,0);
\draw[color=black] (A) to [bend left=45] (B);
\coordinate (A) at (0.75,-0.35);
\coordinate (B) at (0.5,0);
\draw[color=black] (A) to [bend left=45] (B);
\coordinate (A) at (-0.5,0);
\coordinate (B) at (0,-0.85);
\draw[color=black] (A) to [bend left=-30] (B);
\coordinate (A) at (1,0);
\coordinate (B) at (1.5,0.5);
\draw[color=black] (A) to [bend left=45] (B);
\coordinate (A) at (2,0);
\coordinate (B) at (1.5,0.5);
\draw[color=black] (B) to [bend left=45] (A);
\coordinate (A) at (1,0);
\coordinate (B) at (0,-1);
\draw[color=black] (A) to [bend left=25] (B);
\coordinate (A) at (-0.6,-1.8);
\coordinate (B) at (0,-1);
\draw[color=black] (B) to [bend left=-40] (A);
\coordinate (A) at (-0.6,-1.8);
\coordinate (B) at (0,-2.6);
\draw[color=black] (A) to [bend left=-40] (B);
\coordinate (A) at (1.5,-1);
\coordinate (B) at (2,0);
\draw[color=black] (B) to [bend left=25] (A);
\coordinate (A) at (1.5,-1);
\coordinate (B) at (0.75,-1.5);
\draw[color=black] (B) to [bend left=-5] (A);
\coordinate (A) at (0.4,-1.8);
\coordinate (B) at (0.75,-1.5);
\draw[color=black] (B) to [bend left=-30] (A);
\coordinate (A) at (0.4,-1.8);
\coordinate (B) at (0.75,-2.2);
\draw[color=black] (A) to [bend left=-40] (B);
\coordinate (A) at (1.15,-1.4);
\coordinate (B) at (1.5,-1.8);
\draw[color=black] (A) to [bend left=40] (B);
\coordinate (A) at (1.5,-1.8);
\coordinate (B) at (1.2,-2.5);
\draw[color=black] (A) to [bend left=25] (B);
\coordinate (A) at (1.2,-2.5);
\coordinate (B) at (0.8,-2.8);
\draw[color=black] (A) to [bend left=10] (B);
\coordinate (A) at (0,-3.3);
\coordinate (B) at (0.8,-2.8);
\draw[color=black] (B) to [bend left=3] (A);
\coordinate (A) at (0,-3.3);
\coordinate (B) at (-0.6,-4);
\draw[color=black] (A) to [bend left=-30] (B);
\coordinate (A) at (1.85,-0.75);
\coordinate (B) at (2.25,-1.2);
\draw[color=black] (A) to [bend left=35] (B);
\coordinate (A) at (1.85,-2.85);
\coordinate (B) at (2.25,-1.2);
\draw[color=black] (B) to [bend left=35] (A);
\coordinate (A) at (1.85,-2.85);
\coordinate (B) at (1.5,-3.2);
\draw[color=black] (A) to [bend left=10] (B);
\coordinate (A) at (0.2,-4);
\coordinate (B) at (1.5,-3.2);
\draw[color=black] (B) to [bend left=2] (A);
\coordinate (A) at (0.5,-2.8);
\coordinate (B) at (0,-2.6);
\draw[color=black] (B) to [bend left=-10] (A);
\coordinate (A) at (1.12,-2.3);
\coordinate (B) at (0.75,-2.2);
\draw[color=black] (B) to [bend left=-5] (A);
\coordinate (A) at (1.75,-3.2);
\coordinate (B) at (2.2,-4);
\draw[color=black] (A) to [bend left=20] (B);
\coordinate (A) at (2.2,-2.6);
\coordinate (B) at (3,-4);
\draw[color=black] (A) to [bend left=25] (B);
\end{tikzpicture}
\end{center}
\begin{center}
\begin{tikzpicture}
\node (3.75,0.35) at (-3,-1) {$=$};
\node  (3.75,0.35) at (-0.7,0.25) {$j$};
\node  (3.75,0.35) at (0.85,0.08) {$-i$};
\coordinate (A) at (0,0.5);
\coordinate (B) at (-0.5,0);
\draw[color=black] (B) to [bend left=45] (A);
\coordinate (A) at (-0.4,-0.5);
\coordinate (B) at (-0.5,0);
\draw[color=black] (B) to [bend left=-15] (A);
\coordinate (A) at (-0.4,-0.5);
\coordinate (B) at (-0.25,-0.75);
\draw[color=black] (B) to [bend left=5] (A);
\coordinate (A) at (0,0.5);
\coordinate (B) at (0.5,0);
\draw[color=black] (A) to [bend left=45] (B);
\coordinate (A) at (0.6,-0.2);
\coordinate (B) at (0.5,0);
\draw[color=black] (B) to [bend left=-15] (A);
\coordinate (A) at (0.6,-0.35);
\coordinate (B) at (1.3,-0.1);
\draw[color=black] (B) to [bend left=-30] (A);
\coordinate (A) at (0.6,-0.35);
\coordinate (B) at (-0.6,-1.2);
\draw[color=black] (B) to [bend left=-5] (A);
\coordinate (A) at (1.6,-0.3);
\coordinate (B) at (1.3,-0.1);
\draw[color=black] (B) to [bend left=20] (A);
\coordinate (A) at (1.6,-0.3);
\coordinate (B) at (1.6,-0.8);
\draw[color=black] (B) to [bend left=-40] (A);
\coordinate (A) at (1.6,-0.8);
\coordinate (B) at (0,-2);
\draw[color=black] (A) to [bend left=5] (B);
\coordinate (A) at (1.4,-1.1);
\coordinate (B) at (2.4,-1.8);
\draw[color=black] (A) to [bend left=-5] (B);
\coordinate (A) at (0.6,-1.8);
\coordinate (B) at (1.4,-2.4);
\draw[color=black] (A) to [bend left=-5] (B);
\end{tikzpicture}
\end{center}
\begin{center}
\begin{tikzpicture}
\node (3.75,0.35) at (-2.5,-1) {$=q^{i j}$};
\node (3.75,0.35) at (-0.7,-0.6) {$j$};
\node (3.75,0.35) at (0.6,0.05) {$-i$};
\coordinate (A) at (0.6,-0.25);
\coordinate (B) at (1.7,-0.15);
\draw[color=black] (B) to [bend left=-40] (A);
\coordinate (A) at (0.6,-0.25);
\coordinate (B) at (-0.6,-1.2);
\draw[color=black] (B) to [bend left=-5] (A);
\draw (1.7,-0.15) --(1.8,-0.3);
\coordinate (A) at (1.8,-0.3);
\coordinate (B) at (1.6,-0.8);
\draw[color=black] (B) to [bend left=-40] (A);
\coordinate (A) at (1.6,-0.8);
\coordinate (B) at (0,-2);
\draw[color=black] (A) to [bend left=5] (B);
\coordinate (A) at (1.4,-1.1);
\coordinate (B) at (2.8,-1.4);
\draw[color=black] (A) to [bend left=-5] (B);
\coordinate (A) at (0.6,-1.75);
\coordinate (B) at (2.4,-2.1);
\draw[color=black] (A) to [bend left=-5] (B);
\coordinate (A) at (-0.45,-0.6);
\coordinate (B) at (-0.45,-0.95);
\draw[color=black] (B) to [bend left=35] (A);
\coordinate (A) at (-0.3,-0.5);
\coordinate (B) at (-0.45,-0.6);
\draw[color=black] (B) to [bend left=25] (A);
\coordinate (A) at (-0.3,-0.5);
\coordinate (B) at (0.2,-0.52);
\draw[color=black] (A) to [bend left=25] (B);
\end{tikzpicture}
\end{center}
\begin{center}
\begin{tikzpicture}
\node (3.75,0.35) at (-1.8,0.3) {$=q^{i j}$};
\node  (3.75,0.35) at (-0.5,0.6) {$-i$};
\coordinate (A) at (-0.3,0);
\coordinate (B) at (0.45,1);
\draw[color=black] (A) to [bend left=45] (B);
\draw (0.45,1) --(0.55,1);
\coordinate (A) at (0.55,1);
\coordinate (B) at (1.3,0);
\draw[color=black] (A) to [bend left=45] (B);
\coordinate (A) at (2,0);
\coordinate (B) at (2.65,1);
\draw[color=black] (A) to [bend left=45] (B);
\draw (2.65,1) --(2.75,1);
\coordinate (A) at (2.75,1);
\coordinate (B) at (3.5,0);
\draw[color=black] (A) to [bend left=45] (B);
\node  (3.75,0.35) at (1.8,0.2) {$j$};
\end{tikzpicture}
\end{center}
\begin{center}
\begin{tikzpicture}
\node (3.75,0.35) at (-1.7,0.3) {$=q^{2 i j}$};
\node  (3.75,0.35) at (-0.6,0.2) {$-i$};
\coordinate (A) at (-0.3,0);
\coordinate (B) at (0.45,1);
\draw[color=black] (A) to [bend left=45] (B);
\draw (0.45,1) --(0.55,1);
\coordinate (A) at (0.55,1);
\coordinate (B) at (1.3,0);
\draw[color=black] (A) to [bend left=45] (B);
\coordinate (A) at (2,0);
\coordinate (B) at (2.65,1);
\draw[color=black] (A) to [bend left=45] (B);
\draw (2.65,1) --(2.75,1);
\coordinate (A) at (2.75,1);
\coordinate (B) at (3.5,0);
\draw[color=black] (A) to [bend left=45] (B);
\node  (3.75,0.35) at (1.9,0.6) {$j$};
\end{tikzpicture}
\end{center}
\caption{Double Braid Relation.\label{DoubleBraidRelation}}
\end{figure}
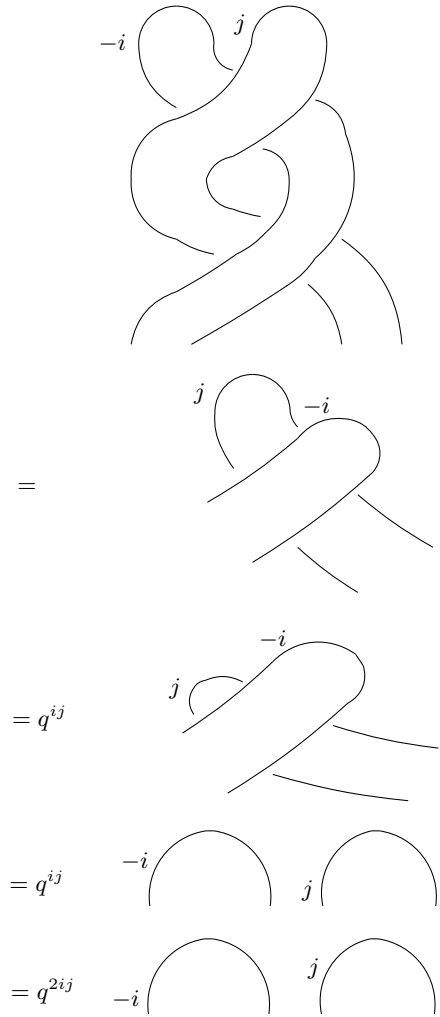

The double braid $S$ is illustrated in Figure \ref{Double Braid}.
\noindent It preserves 2-qudits.  Furthermore the double braid is the square of the controlled $Z$, namely  
	\be
	S=C_Z^2\;.
	\label{DoubleBraidCZ}
	\ee
The relation \eqref{DoubleBraidCZ} has been shown to be true by using algebraic identities~\cite{HutterLoss}. 
Here we give an elementary proof using diagrams.  In fact the proof follows from the qudit-braid relation given in Figure \ref{Braid-qudit}. We illustrate our proof with the isotopy in Figure \ref{DoubleBraidRelation}. 

One can obtain qudit matrices $X$, $Y$ from the matrix $Z$ by the conjugation of braids $b_{i}$ of the $i^{\textrm{th}}$-string given in  Figure \ref{Braid-bi}.  As in \S5.2.2 of \cite{JL}, one has:
\begin{align*}
Y&=b_2 Z^* b_2^{*};\\
X&=b_1b_2 Z b_2^{*}b_1^{*}.
\end{align*}
Thus
\begin{align*}
C_Y^2&=b_6 S^* b_6^{*};\\
C_X^2&=b_5b_6 S b_6^{*}b_5^{*}.
\end{align*}
Correspondingly, both $C_Y^2$ and $C_X^2$ also preserve the subspace spanned by qudits and both are represented by braided diagrams.

\section{Entanglement-Relay Protocol}\label{Sect:ERelay}
One can extend our earlier discussion in \S\ref{sec:ESwap} in order to create an entanglement-relay network to produce and share  \textit{long-distance entanglement}. We now show how to enable entanglement that is non-local. Such entanglement might be across a device, allowing for non-local entangling gates, or it might involve a network of distributed devices. 

Let us describe the situation in detail in the case that Alice wants to entangle her qudit with the qudit of Bob, utilizing the aid of three intermediate helpers, H$1$, H$2$, and H$3$, who line up from left to right. The idea is that each person uses a nearest-neighbor for entanglement swapping. Each swap involves one measurement station, which we might call M$1$, \ldots, M$4$. We illustrate this protocol in Figure \ref{ThreeHelpers}.   
\begin{center}
\begin{figure}[h]
\scalebox{0.66}{
\begin{tikzpicture}
\node  (3.75,0.35) at (1.3,3.2) {$\mbox{Alice}$};
\node  (2.75,0.35) at (11.7,2.55) {$\mbox{Alice}$};
\node  (2.75,0.35) at (13.05,2.55) {$\mbox{Bob}$};
\node  (3.75,0.35) at (3.5,3.2) {$\mbox{H}1$};
\node  (3.75,0.35) at (5.5,3.2) {$\mbox{H}2$};
\node  (3.75,0.35) at (7.5,3.2) {$\mbox{H}3$};
\node  (3.75,0.35) at (8.9,3.2) {$\mbox{Bob}$};
\node  (3.75,0.35) at (1.9,0.4) {$\mbox{M}1$};
\node  (3.75,0.35) at (3.9,0.4) {$\mbox{M}2$};
\node  (3.75,0.35) at (5.9,0.4) {$\mbox{M}3$};
\node  (3.75,0.35) at (7.9,0.4) {$\mbox{M}4$};
\draw[->] (10.2,1.8) --(11,1.8);
\draw (12.35,1.9) --(12,2.25);
\draw (13,1.25) --(12.65,1.6);
\draw (12,1.25) --(13,2.25);
\draw (1.25,2.5) --(1,2.75);
\draw (2.25,1.5) --(1.5,2.25);
\draw (1,2) --(1.75,2.75);
\draw (2.15,1.8) --(4,2.75);
\draw (1.5,1.5) --(1.95,1.7);
\coordinate (A) at (1,2);
\coordinate (B) at (0.8,1.6);
\draw[color=black] (A) to [bend left=-20] (B);
\draw (0.8,0.4) --(0.8,1.6);
\draw (2.25,1.5) --(1.5,1.5);
\draw (1.5,1.5) --(1.5,0.8);
\draw (2.25,1.5) --(2.25,0.8);
\draw (2.25,0.8) --(1.5,0.8);
\draw (2.15,1) arc (0:180:.25);
\draw (1.85,1) --(2.1,1.4);
\draw (3.25,2.5) --(3,2.75);
\draw (4.25,1.5) --(3.5,2.25);
\draw (3.5,1.5) --(3.95,1.7);
\draw (4.25,1.5) --(3.5,1.5);
\draw (3.5,1.5) --(3.5,0.8);
\draw (4.25,1.5) --(4.25,0.8);
\draw (4.25,0.8) --(3.5,0.8);
\draw (4.15,1) arc (0:180:.25);
\draw (3.85,1) --(4.1,1.4);
\draw (4.15,1.8) --(6,2.75);
\draw (5.25,2.5) --(5,2.75);
\draw (6.25,1.5) --(5.5,2.25);
\draw (5.5,1.5) --(5.95,1.7);
\draw (6.25,1.5) --(5.5,1.5);
\draw (5.5,1.5) --(5.5,0.8);
\draw (6.25,1.5) --(6.25,0.8);
\draw (6.25,0.8) --(5.5,0.8);
\draw (6.15,1) arc (0:180:.25);
\draw (5.85,1) --(6.1,1.4);
\draw (6.15,1.8) --(8,2.75);
\draw (7.25,2.5) --(7,2.75);
\draw (8.25,1.5) --(7.5,2.25);
\draw (8.25,1.5) --(7.5,1.5);
\draw (7.5,1.5) --(7.5,0.8);
\draw (8.25,1.5) --(8.25,0.8);
\draw (8.25,0.8) --(7.5,0.8);
\draw (8.15,1) arc (0:180:.25);
\draw (7.85,1) --(8.1,1.4);
\draw (7.5,1.5) --(7.85,1.78);
\draw (8,1.87) --(9.25,2.75);
\draw (8.75,2.5) --(8.5,2.75);
\draw (9.25,2) --(8.9,2.35);
\coordinate (A) at (9.25,2);
\coordinate (B) at (9.45,1.6);
\draw[color=black] (A) to [bend left=20] (B);
\draw (9.45,0.4) --(9.45,1.6);
\end{tikzpicture}}
\caption{Entanglement-Relay with Three Helpers. \label{ThreeHelpers}
}
\end{figure}
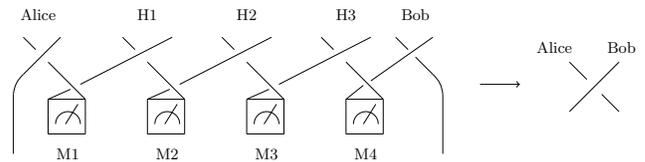
\end{center}
The resolution of the protocol implements topological isotopy. This produces an overall constant $\delta^{-8}$ which we can ignore, as this factor does not affect the entanglement.

The end result is the maximal entanglement of Alice's qudit with the qudit of Bob. It is clear that this situation generalizes for any number  $\#H$ of helpers, in which case the constant would be~$\delta^{-2(\#H+1)}$.

\begin{acknowledgements}
\noindent This research was supported in part by a grant from the Templeton Religion Trust.  We are also grateful for hospitality at the FIM of the ETH-Zurich, where part of this work was carried out.
\end{acknowledgements}

\end{document}